\documentclass[twocolumn]{aastex631}
\usepackage{graphicx}
\usepackage[T1]{fontenc}
\usepackage{amssymb}
\usepackage{amsmath}

\begin{document}

\title{Gravitational instability, spiral substructure, and modest grain growth in a typical protostellar disk:\\
Modeling multi-wavelength dust continuum observation of TMC1A}

\author[0000-0002-9408-2857]{Wenrui Xu}
\affiliation{Center for Computational Astrophysics, Flatiron Institute, New York, USA}

\author[0000-0002-9661-7958]{Satoshi Ohashi}
\affiliation{National Astronomical Observatory of Japan, Tokyo, Japan}

\author[0000-0002-8238-7709]{Yusuke Aso}
\affiliation{Korea Astronomy and Space Science Institute, Daejeon, Republic of Korea}

\author[0000-0003-2300-2626]{Hauyu Baobab Liu}
\affiliation{Physics Department, National Sun Yat-Sen University, Taiwan, Republic of China}

\begin{abstract}
Embedded, Class 0/I protostellar disks represent the initial condition for planet formation. This calls for better understandings of their bulk properties and the dust grains within them.
We model multi-wavelength dust continuum observations of the disk surrounding the Class I protostar TMC1A to provide insight on these properties.
The observations can be well fit by a gravitationally self-regulated (i.e., marginally gravitationally unstable and internally heated) disk model, with surface density $\Sigma \sim 1720 (R/10{\rm au})^{-1.96}~{\rm g~cm}^{-2}$ and midplane temperature $T_{\rm mid} \sim 185 (R/10{\rm au})^{-1.27}~{\rm K}$.
The observed disk contains a $m=1$ spiral substructure; we use our model to predict the spiral's pitch angle and the prediction is consistent with the observations.
This agreement serves as both a test of our model and strong evidence of the gravitational nature of the spiral.
Our model estimates a maximum grain size $a_{\rm max}\sim 196(R/10\rm au)^{-2.45} \mu{\rm m}$, which is consistent with grain growth being capped by a fragmentation barrier with threshold velocity $\sim1~{\rm m/s}$.
We further demonstrate that observational properties of TMC1A are typical among the observed population of Class 0/I disks, which hints that traditional methods of disk data analyses based on Gaussian fitting and the assumption of the optically thin dust emission could have systematically underestimated disk size and mass and overestimated grain size.
\end{abstract}


\section{Introduction}\label{sec:intro}
\nocite{XK21a} 

When and how does planet formation begin in young stellar systems?
Substructures that might have formed via planet-disk interaction are ubiquitously detected in young protoplanetary disks \citep{dsharpI,dsharpII,dsharpVII,Andrews2020}, suggesting that planet formation probably begins very early, and the first stages of planet formation -- the growth of dust grains into pebbles and/or planetesimals -- are probably already underway during the main accretion phase (Class 0/I). Recent detections of disk substructures in a number of Class 0/I disks \citep{Nakatani2020,Sheehan2020,Segura-Cox2020,Ohashi2022} also support this idea, with the caveat that it is often less clear whether these substructures are related to planets.

Observationally constraining the first stages of planet formation in Class 0/I disks, however, remains challenging.
In particular, it is difficult to constrain the dust mass and dust grain size reliably.
The common practice of using the dust continuum emission to estimate dust mass from flux density and constrain grain size with spectral index requires the dust emission in the disk to be optically thin at the observed wavelength.
However, Class 0/I disks (as well as some young Class II disks) probably have higher dust surface density compared to older Class II disks, causing them to be optically thick at $\lesssim$mm wavelength \citep{Li2017,Galvan-Madrid2018,Tobin2020,Ko2020ApJ...889..172K,Liu2021,Zamponi2021}.
When the disk is optically thick, it would be difficult to tell how much dust mass remains invisible (at high optical depth) and whether a low spectral index is due to large grain size \citep{Draine2006} or high optical depth.
Additionally, radial variation of disk properties (surface density, grain size) makes it challenging to constrain parameters of a sufficiently generic disk model without significant degeneracy, especially when resolution is limited \citep[e.g.,][Section 6.4]{Tazzari2021tension}.
Another concern is the temperature profile of the disk; traditional models generally assume that the disk is passively heated by the protostar (following models of older Class II disks, cf. \citealt{ChiangGoldreich1997}), yet several theoretical and observational studies \citep[e.g.,][]{XK21b,Zamponi2021,Xu22} suggest that the high accretion rate in Class 0/I disks, together with self-shielding of protostellar irradiation, makes internal viscous heating the dominate heating source.

In this study, we address these concerns with a case study of the Class I disk TMC1A. TMC1A is an ideal subject for such a study, given existing high-quality data and independent estimates on dust properties in the disk from different techniques \citep{Harsono2018,Aso2021}.
More importantly, the disk around TMC1A is likely a representative sample of Class 0/I disks because its observational properties are typical for Class 0/I systems (Section \ref{sec:typical}).

The rest of this paper is organized as follows.
In Section \ref{sec:model} we describe our data and model.
We then discuss the results of our modeling on surface density and temperature (Section \ref{sec:results}), spiral substructure (Section \ref{sec:spiral}), and dust grain size (Section \ref{sec:dust}).
We discuss broader implications of our results in Section \ref{sec:discussion} and summarize our conclusions in Section \ref{sec:conclusion}.

\section{Modeling multi-wavelength observations of TMC1A}\label{sec:model}

\begin{figure*}
    \centering
    \includegraphics[scale=0.6]{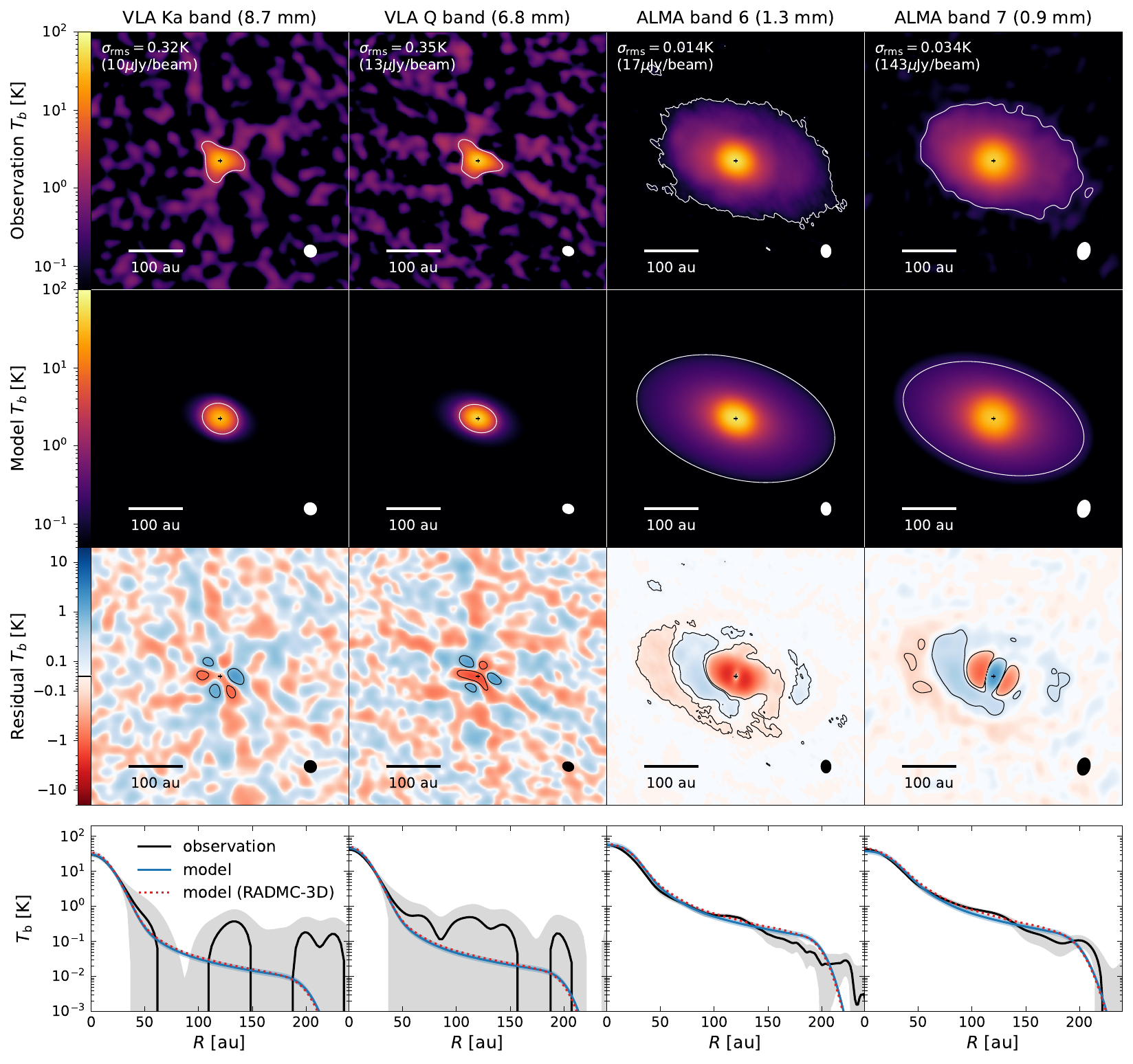}
    \caption{Comparison between observations and our model (with fiducal parameters) at different wavelengths.
    The first row shows the observed brightness temperature ($T_{\rm b}$) maps; the second row shows the mock observations produced by our axisymmetric disk model; the third row shows the residual maps (observation$-$model).
    We label the disk center (black crosses), the RMS noise $\sigma_{\rm rms}$, and the $5\sigma_{\rm rms}$ contours (white for the first two rows, black for the third row) for reference.
    The last row shows cuts along the major axis; there the observation (black lines) has been averaged between the two sides of the major axis, and shaded areas show model and observational uncertainties (cf. Appendix \ref{a:fit}).
    Our model agrees relatively well with observation at all four wavelengths, correctly capturing the radial and wavelength dependences of the brightness temperature profile. Note that we only model the axisymmetric component of the emission here; the asymmetry in the outer part of the ALMA images is mainly the result of an $m=1$ spiral substructure, which we discuss in Section \ref{sec:spiral} and Fig. \ref{fig:spiral}. We also include results from a full radiative transfer calculation (red dotted lines) to demonstrate that ignoring protostellar irradiation in our model does not cause significant error.}
    \label{fig:obs_vs_model}
\end{figure*}

\subsection{Data}
We model ALMA and VLA dust continuum observations at four different wavelengths, 0.9, 1.3, 6.8 and 8.7~mm (Fig. \ref{fig:obs_vs_model}, first row). The 0.9 and 1.3~mm images combine several archival ALMA data sets (\citealt{Harsono2018,Harsono2021,Aso2021}; project code 2015.1.01415.S, 2015.1.01549.S, and 2018.1.00701.S), and the 6.8 and 8.7~mm images are produced from new VLA observations (project code 21B-089). The VLA observations and ALMA data reprocessing are detailed in Aso et al. (in prep). All images are reconstructed using Briggs weighting with a robust parameter of 0.5, yielding a synthesized beam size of $0\farcs17$ -- $0\farcs25$ (24 -- 36~au). In Fig. \ref{fig:obs_vs_model} we show the synthesized beam and root-mean-squared (RMS) noise of each image. The location of the protostar (marked by black crosses in Fig. \ref{fig:obs_vs_model}) is determined by fitting a 2D Gaussian profile to the observed image at each wavelength. We expect this to yield a reasonably good estimate despite the presence of non-axisymmetric perturbation (mainly a $m=1$ spiral; see Section \ref{sec:spiral}) in the disk because the Gaussian fit is mainly determined by the inner part of the disk, which contains most of the flux (cf. Section \ref{sec:apparent}) and is less affected by the spiral substructure compared to the outer disk (cf. Section \ref{sec:spiral_amplitude}). We estimate the protostar location independently at each wavelength, because the observations at different wavelengths were conducted several years apart and the proper motion of the protostar might shift its sky coordinates.

Additionally, we make use of the dynamical mass constraint from the CO velocity measurement in \citet{Aso2015},
\begin{equation}
     M_{\rm p} \equiv M\sin^2 i = 0.56\pm 0.05 M_\odot.\label{eq:Mp}
\end{equation}
Here we use $M\sin^2 i$ instead of $M$ because the former is directly related to the observed line-of-sight velocity.
In our fitting we use the mass $M$ above to constrain the total mass of the star and the disk $M_{\rm tot}$. 
In our disk model, the disk has a steep surface density profile and keeps most of the mass at small radii (cf. Section \ref{sec:results_overview}); therefore at most radii the rotation is close to the Keplerian rotation with the central mass $M_{\rm tot}$.

\subsection{Disk model}\label{sec:model_summary}
We use a parameterized semi-analytic model of protostellar disks to generate azimuthally averaged disk properties and multi-wavelength mock observations. Because this model is axisymmetric (with turbulent fluctuations, including sprials, modeled as an effective viscosity), the mock observations only include the axisymmetric component of the emission and do not include asymmetric substructures such as the spiral discussed in Section \ref{sec:spiral}.
This model has been detailed in \citet{Xu22}; below we summarize the assumptions and features of this model and a few minor modifications.

The central assumption of our model is that the disk is gravitationally self-regulated; i.e., the effective viscosity produced by gravitational instability (GI) serves as the main mechanism for both angular-momentum transport and heating \citep{XK21a,XK21b}. This translates to two more specific assumptions: First, the temperature profile is set by an internal heating rate $\propto\dot M$; second, the disk is marginally unstable (Toomre $Q=1$--2). Here we only assume the disk to be marginally unstable without further constraining $Q$ because the effective viscosity of GI $\alpha_{\rm GI}$ cannot be parametrized accurately as a function of local disk properties (nor is the result sensitive to the exact relation between $\alpha_{\rm GI}$ and $Q$; see discussions in \citealt{Zhu2010}). This choice is similar to the constant-$Q$ closure discussed in \citet{Rafikov2015}. We comment on physical and observational motivations of these assumptions in Section \ref{sec:assumption}.

Under these assumptions, the azimuthally averaged radial profile of the surface density and the radial and vertical profile of temperature can be solved from the following parameters (see method of solution in \citealt{Xu22}): 
total mass of the star and the disk $M_{\rm tot}$, disk size $R_{\rm d}$, accretion rate $\dot M$ (assumed to be constant throughout the disk for simplicity), Toomre $Q$ parameter, maximum grain size $a_{\rm max}$, and the power-law index of the grain size distribution $q$ (with ${\rm d}n/{\rm d}a \propto a^{-q}$).
Here the grain size distribution determines the opacity of the disk; we adopt the DSHARP opacity model for opacity computation (\citealt{dsharpV}; see examples of the opacity as a function of $a_{\rm max}$ in their Fig. 4 or \citealt{Xu22} Fig. 1).
There is no free parameter that directly controls the radial slopes of surface density and temperature; both are determined self-consistently by the requirements of thermal equilibrium and marginal GI.
Our model also does not include the magnetic field, because the assumption of gravitational self-regulation implies that the density and temperature profiles are insensitive to the magnetic field inside the disk (cf. Section \ref{sec:assumption}).

\citet{Xu22} fix Toomre $Q$, $a_{\rm max}$, and $q$ to fiducial values. For this work, since we have better observational constraints, we allow Toomre $Q$ and $a_{\rm max}$ to take arbitrary power-law radial profiles specified by their values at 10 and 100~au ($Q^{10{\rm au}}, Q^{100{\rm au}}, a_{\rm max}^{10{\rm au}}, a_{\rm max}^{100{\rm au}}$), and leave $q$ as a free parameter;
in our fiducial model we also assume $Q$ to be marginal throughout the disk (see next subsection).

We comment that our model does not include an envelope component.
This is because for our observations the envelope emission is probably much weaker than the disk emission and could be below the detection limit. The detected dust emission at mm wavelengths show a disk-like morphology, whereas the envelope is expected to be more asymmetric and filamentary \citep[cf.][]{Pineda2022}.
Previous analysis by \citet{Aso2015} also found that CO emission from the innermost 200~au can be fit well with a single Keplerian component, suggesting that envelope contamination is probably negligible.

Our model produces multi-wavelength mock observations by solving radiative transfer under thin disk approximation; the calculation has been detailed in \citet{Xu22} Section 3.2. The potentially important effect of scattering \citep[cf.][]{Liu2019,Zhu2019} has been incorporated in the calculation. Meanwhile, our model does not include the heating from protostellar irradiation, because disk self-shielding makes stellar heating inefficient (cf. Section \ref{sec:assumption}). We adopt these simplifications (thin disk, no stellar heating) mainly to keep the computational cost of our MCMC fitting (Section \ref{sec:fitting}) manageable. We also compare our results with a more realistic radiative transfer calculation in Section \ref{sec:test} to demonstrate that these simplifications do not lead to large errors.

\subsection{Fitting model to observation}\label{sec:fitting}
To constrain the free parameters of our model (disk parameters defined in Section \ref{sec:model_summary}, including $M_{\rm tot}$, $R_{\rm d}$, $\dot M$, $Q^{10{\rm au}}$, $Q^{100{\rm au}}$, $a_{\rm max}^{10{\rm au}}$, $a_{\rm max}^{100{\rm au}}$, $q$, together with inclination $i$ and position angle $\theta$) we perform a MCMC fit using the \verb|emcee| package \citep{Foreman-Mackey2013}. We use broad, uninformative priors for all variables, with the exceptions of $Q$ and $q$. For the dust size distribution $q$, we choose a uniform prior of $[2.5,3.5]$. For Toomre $Q$, we adopt the assumption of marginal instability and only consider $Q^{10{\rm au}}$, $Q^{100{\rm au}}$ values satisfying the requirement that $Q\in[1,2]$ everywhere between 5 au and $R_{\rm d}$. We also fit another model without this assumption for testing; see Section \ref{sec:test}.
The log-likelihood function incorporate observations at all four wavelengths as well as the dynamical mass constraint. Other details of the MCMC fitting are discussed in Appendix \ref{a:fit}.

\subsection{Motivations of model assumptions}\label{sec:assumption}
The physical idea of gravitational self-regulation -- that angular-momentum transport by GI leads to a disk that is marginally unstable everywhere -- is originated from \citet{VorobyovBasu2007}, and similar ideas have been discussed or adopted in a number of studies \citep[e.g.,][]{LinPringle1987,Gammie2001,Zhu2010,Rafikov2015}. It later became less popular because magnetic field is believed to play a more important role in disk transport \citep[cf.][]{Lesur2022,Tsukamoto2022} and the complex dynamics of GI makes people question whether it can be treated as a simple viscosity \citep{LodatoRice2004,Cossins2009}.
The low mm fluxes (and thus low apparent masses) of disks are also often interpreted as disfavoring gravitational instability.

However, \citet{XK21a,XK21b} recently found that Class 0/I disks in radiative non-ideal MHD simulations can be well-described by a gravitationally self-regulated model that does not include the magnetic field. While the magnetic field plays an important role in setting the total angular momentum of the disk during the collapse of the pre-stellar core, strong ambipolar diffusion due to low ionization in the disk leaves GI the dominant source of angular-momentum transport within the disk, and the magnetic field is dynamically unimportant within the disk. Meanwhile, despite the complex dynamics of GI, its transport and heating can be approximated as a local viscosity to within some $\mathcal O(1)$ factor.
Later on, \citet{Xu22} further demonstrates that the majority of Class 0/I disks in a recent multi-wavelength survey are consistent with being gravitationally self-regulated, and the low apparent mass is mainly a result of high optical depth hiding most disk mass. High masses in young disks have also been independently demonstrated in observational studies using techniques that are less affected by high optical depth \citep[e.g.,][]{McClure2016,Terry2022,Lodato2023}.

These previous works serve as the basis of the assumptions in this paper. Still, we stress that we do not take these assumptions for granted. Instead, we test them by relaxing these assumptions in Section \ref{sec:test} and by comparing against another independent observational constraint in Section \ref{sec:spiral}.

\section{Surface density and temperature: gravitational self-regulation}\label{sec:results}

\begin{figure*}
    \centering
    \includegraphics[scale=0.6]{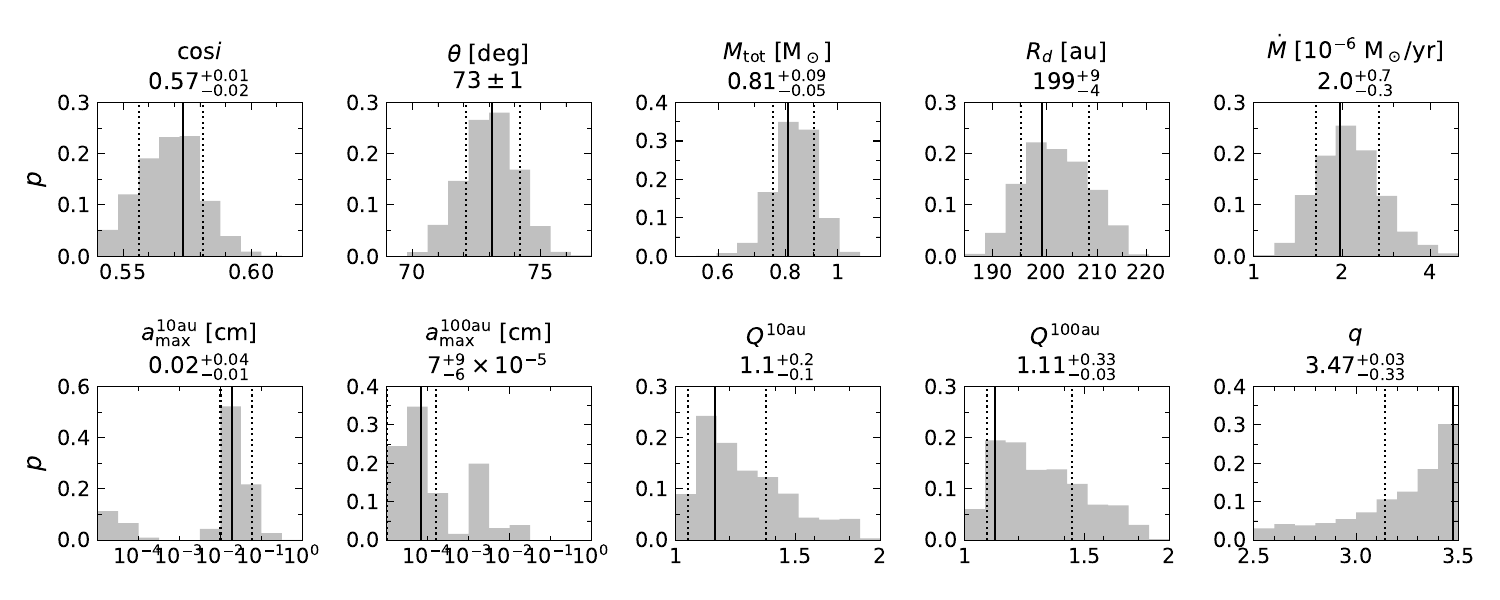}
    \includegraphics[scale=0.6]{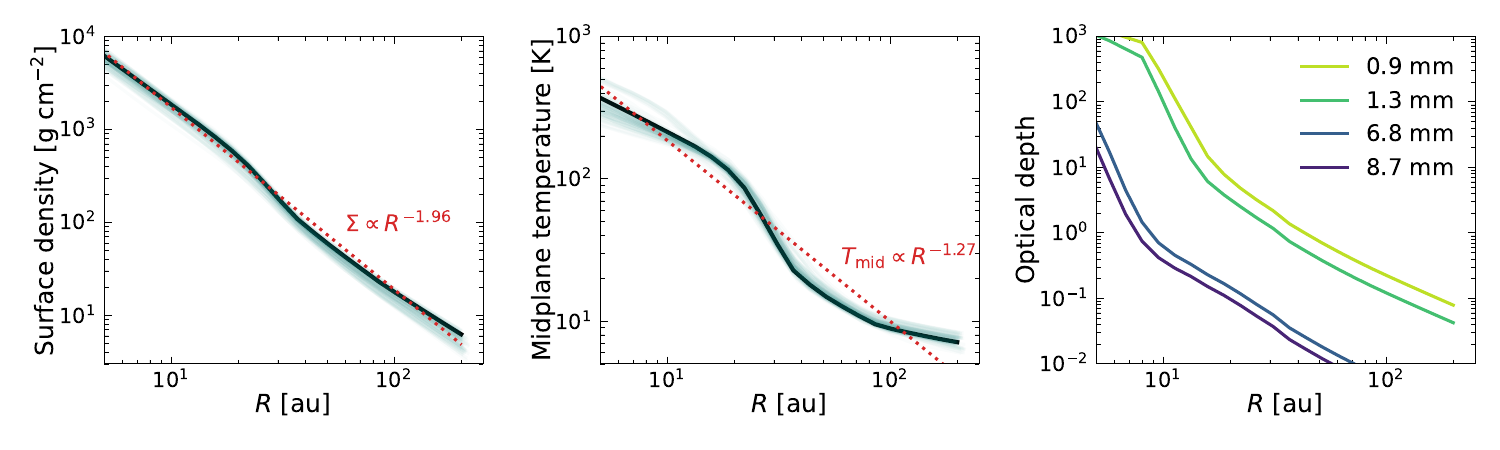}
    \caption{Top panels: a summary of our posteriors (see definitions of parameters in Section \ref{sec:model_summary} and \ref{sec:fitting}). The vertical axes show the probability of each bin. Solid and dashed lines mark the fiducial value (peak of posterior distribution) and $1\sigma$ (68\%) confidence interval, respectively. Most parameters are relatively well constrained.
    Bottom panels: Radial profiles of surface density $\Sigma$ (left), midplane temperature $T_{\rm mid}$ (middle), and optical depth for each observed wavelength (right) for our fiducial disk parameters. We caution that we cannot determine accurately the inner radius where these profiles remain valid ($R_{\rm in}$), because varying $R_{\rm in}$ barely affects the dust continuum observation when $R_{\rm in}$ remains unresolved and optically thick at all observed wavelengths. Light green lines in the left and center panels correspond to disk parameters sampled from the posterior distribution. We also show power-law fits of the radial profiles of $\Sigma$ and $T_{\rm mid}$ for reference.
    }
    \label{fig:model_summary}
\end{figure*}

\begin{figure}
    \centering
    \includegraphics[scale=0.6]{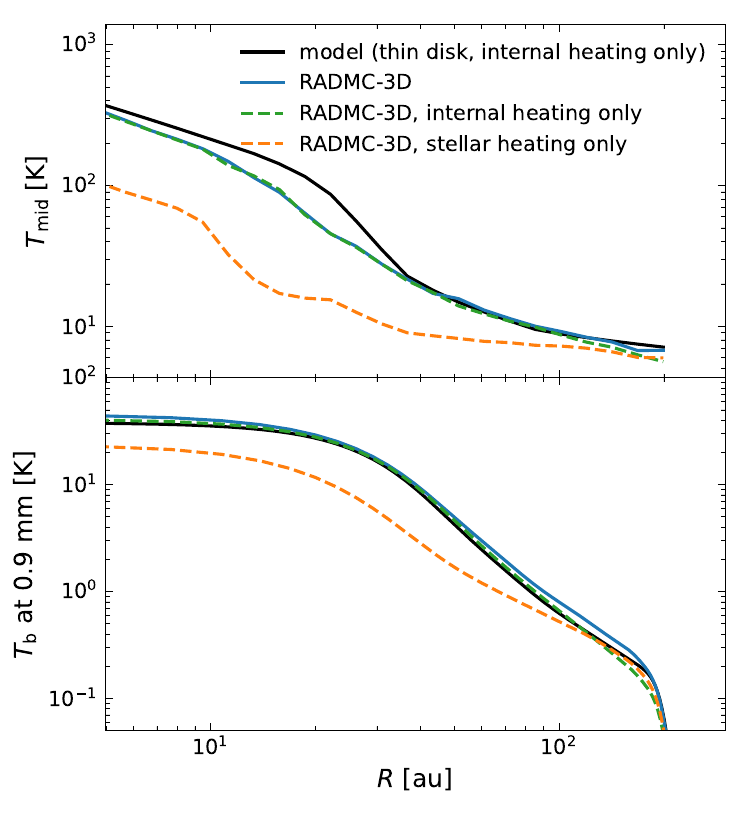}
    \caption{Comparison of midplane temperature (top panel) and brightness temperature at 0.9 mm (bottom panel; other wavelengths show similar results) between our model \citep[cf.][]{Xu22} and \texttt{RADMC-3D} radiative transfer calculations. The radiative transfer results are overall similar to our model. The temperature and emission of the disk are mainly determined by internal heating as opposed to protostellar irradiation.}
    \label{fig:sup_radmc}
\end{figure}

\subsection{Overview}\label{sec:results_overview}
Our model fits well to observations at all four wavelengths (Fig. \ref{fig:obs_vs_model}). The surface density $\Sigma$ and midplane temperature $T_{\rm mid}$ profiles of the disk can be relatively well constrained, giving a warm and massive disk that is optically thick for $\lesssim$mm wavelengths at $R\lesssim 40$~au, as summarized in Fig. \ref{fig:model_summary}.
The radial profiles of $\Sigma$ and $T_{\rm mid}$ in our fiducial model can be roughly described by the following power-laws:
\begin{align}
    &\Sigma \sim 1720 (R/10{\rm au})^{-1.96}~{\rm g~cm}^{-2},\label{eq:Sigma}\\
    &T_{\rm mid} \sim 185 (R/10{\rm au})^{-1.27}~{\rm K}.\label{eq:Tmid}
\end{align}
Both profiles show relatively steep radial slopes, which is a generic property of gravitationally self-regulated disks (cf. \citealt{Xu22} Section 6.1).
The $\Sigma$ and $T_{\rm mid}$ profiles we obtain are not simple power-laws even though our model assumes power-law $Q$ and $a_{\rm max}$ profiles, and this may be primarily due to the complex temperature dependence of the dust opacity (see an example in \citealt{Xu22} Fig. 1).

The total mass of the disk and the star in our model is $M_{\rm tot} = 0.81~M_\odot$. This is significantly larger than the central mass of $0.68~M_\odot$ reported by \citet{Aso2015}. The difference is mainly because the model in \citet{Aso2015} could not accurately determine inclination, resulting in large uncertainty when converting $M\sin^2 i$ (which is directly proportional to line-of-sight velocity) to mass. Meanwhile, our model gives $M\sin^2 i=0.54~M_\odot$, which agrees with the result in \citet[][cf. Eq. \ref{eq:Mp}]{Aso2015}.
However, we are unable to accurately determine what fraction of this mass belongs to the disk, because the steep surface density profile ($\sim$ constant mass per $\log R$) keeps most disk mass at the inner part of the disk and the total disk mass is sensitive to the inner boundary of the gravitationally self-regulated region ($R_{\rm in}$).
In our model we adopt a very conservative estimate of $R_{\rm in}$ following \citet{Xu22}; we do not attempt to constrain $R_{\rm in}$ observationally because varying $R_{\rm in}$ barely affects the observed dust continuum emission when $R_{\rm in}$ remains unresolved and optically thick at all observed wavelengths.
Meanwhile, a useful diagnostic for characterizing the disk mass is the local disk-to-star mass ratio (more precisely, disk-to-total-mass ratio)\footnote{For a less steep surface density profile, $\pi R^2 \Sigma(R)$ evaluated around $R_{\rm d}$ generally differs from the total disk mass only by an order-unity factor ($1-p/2$ for a power-law profile $\Sigma\propto R^{-p}$).}
\begin{equation}
    \frac{\pi R^2 \Sigma(R)}{M_{\rm tot}} \sim 0.075.
\end{equation}
The $\approx -2$ slope of radial density profile makes this ratio insensitive to radius.

\subsection{Testing model assumptions}\label{sec:test}
As a sanity check we perform two simple tests for our model assumptions. First, to test the assumption of marginal instability, we refit our model while relaxing the assumption of marginally unstable $Q$ and instead use broad priors (log uniform between 0.1 and $10^3$) for $Q^{10\rm au}$ and $Q^{100\rm au}$. The resulting estimates of model parameters and disk profiles remain similar to our fiducial model.
The posterior gives $Q^{10\rm au}\sim [0.7,1.4]$ and $Q^{100\rm au}\sim [0.8,1.2]$ (68\% confidence interval).
This demonstrates that (under other assumptions of our model) observation does prefer a disk that is marginally unstable ($Q\sim\mathcal O(1)$) everywhere.

The other key assumption of our model is that the disk is mainly heated internally via the effective viscosity of GI as opposed to external stellar heating.
To test this assumption, we perform radiative transfer calculations on our fiducial disk model using \texttt{RADMC-3D} \citep{Dullemond2012}, which accounts for both internal and external heating. The setup of this calculation is detailed in Appendix \ref{a:radmc}, and the resulting emission profiles are shown in the bottom row of Fig. \ref{fig:obs_vs_model}.
The difference between the \texttt{RADMC-3D} calculation and our model remains relatively small.
We also find that removing protostellar irradiation in our \texttt{RADMC-3D} calculation barely affects the results, but removing internal heating significantly reduces dust temperature and observed emission (Fig. \ref{fig:sup_radmc}).

These tests demonstrate that the assumptions of our model are self-consistent.
We caution that self-consistency alone does not guarantee that the assumptions of a model are applicable to a particular observed system.
The ability to reproduce observational data provides some evidence for applicability, but a major limitation of such evidence is the possibility of overfitting: Due to the potential degeneracy when translating disk properties to observables (cf. Section \ref{sec:intro}) it is possible to reproduce observation by tuning the parameters of an incorrect but sufficiently flexible model.
This limitation can be avoided by testing the model's predictive power, i.e., whether it makes correct predictions on data that are not used for fitting/tuning the model parameters.
Such predictive power has not yet been commonly achieved in the modeling of protostellar/protoplanetary disk observations.
Previously, we have demonstrated our model's ability to predict the systematic decrease of the apparent disk size towards longer wavelengths and the typical spectral index of a population of Class 0/I disks \citep{Xu22}.
The comparison in the next section also serves as a test of our model's predictive power (i.e., the ability to reproduce independent, observed signatures) for TMC1A.

\section{Spiral substructure}\label{sec:spiral}

\subsection{Pitch angle: testing our model and evidence of GI}\label{sec:spiral_pitch}

\begin{figure*}
    \centering
    \includegraphics[scale=0.6]{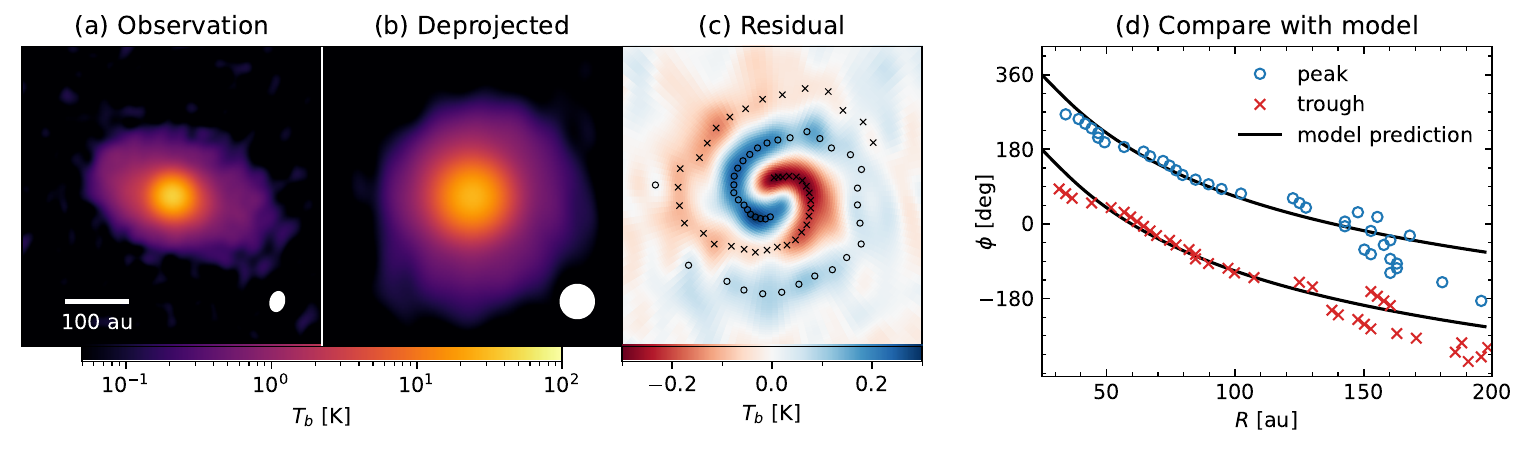}
    \caption{The $m=1$ spiral substructure in TMC1A. (a) Observed $T_{\rm b}$ map at 0.9~mm. (b) Deprojecting the observation to face-on. To ensure that the beam does not introduce asymmetry, we also make the (effective) synthesized beam circular as follows. After the deprojection, which also stretches the synthesized beam, we convolve the image with a 1D Gaussian kernel along the minor axis of the stretched beam. The width of this second Gaussian is chosen so that the combined effect of the stretched synthesized beam and this new Gaussian kernel is equivalent to a circular 2D Gaussian beam. This produces a circular beam with a width (FWHM) of 59~au. (c) Residual $T_{\rm b}$ after subtracting the azimuthal average. Markers show the peak and trough of the $m=1$ spiral, whose locations are defined by local extrema in the radial direction. (d) Comparison between observed spiral geometry (markers) and our model's prediction (black lines). The model prediction is computed from the linear dispersion relation Eq. \ref{eq:kr}, using the radial surface density and sound speed profiles in our model. The good agreement between the two suggests that the spiral is likely excited by GI.
    }
    \label{fig:spiral}
\end{figure*}

\begin{figure}
    \centering
    \includegraphics[scale=0.6]{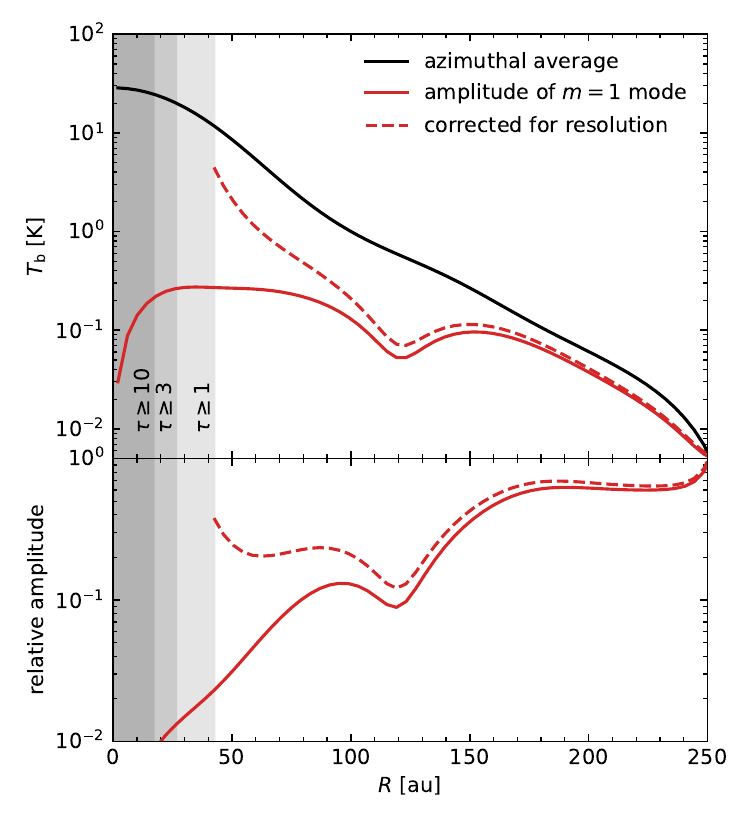}
    \caption{Amplitude of the $m=1$ spiral wave in the deprojected 0.9~mm image (Fig. \ref{fig:spiral} panel b). Grey shades show the region with high line-of-sight optical depth, where the relative amplitude of $T_{\rm b}$ perturbation can be much lower than that of the underlying surface density perturbation. Red dashed lines show the estimated amplitude of $T_{\rm b}$ perturbation after correcting for finite resolution (Section \ref{sec:spiral_amplitude}) in the optically thin region.
    }
    \label{fig:spiral_amplitude}
\end{figure}

Thanks to high-resolution and high-sensitivity observations, TMC1A demonstrates a clear $m=1$ spiral arm at mm wavelengths once we subtract the axisymmetric component of the emission (Fig. \ref{fig:spiral}; also see \citealt{Aso2021}). To check whether this spiral arm is consistent with being excited by GI, we compare the measured geometry (pitch angle) of the spiral with our model's prediction.

An important feature of gravitationally excited spiral arms is that the pitch angle (or radial wavenumber $k_r$) of the most unstable mode is uniquely determined by the surface density and temperature profiles of the disk.
The linear dispersion relation of tightly wound\footnote{The spiral in TMC1A has $k_rr/m\sim 4$, so this is still a reasonable approximation.} spiral density waves in a self-gravitating disk is given by \citep[][Eq. 16]{KratterLodato2016}
\begin{equation}
    (\omega- m \Omega)^2 = (c_{\rm s,eff}^2 + 2\pi G\Sigma H) k_r^2 - 2\pi G \Sigma |k_r| + \kappa^2.\label{eq:spiral_dispersion}
\end{equation}
Here $\omega$ is the pattern frequency, $m$ is the azimuthal wavenumber ($m=1$ in our case), $\Omega$ and $\kappa$ are the rotation rate and epicyclic frequency, $c_{\rm s, eff}$ is the effective sound speed for a thin disk (defined by $c_{\rm s, eff}^2 = \partial P/\partial \Sigma$ where $P$ is the vertically integrated pressure), and $H = c_{\rm s}/\Omega$ is the scale height.\footnote{A more accurate formula of $H$ when self-gravity is non-negligible is given in \citet{BertinLodato1999}. For our disk model, adopting that formula modifies $H$ by $\sim10\%$ and barely affects the final result for $k_r$.}
Following \citet{Goldreich1986}, $c_{\rm s, eff}$ is related to the sound speed $c_{\rm s}$ via $c_{\rm s, eff}^2 = c_{\rm s}^2\gamma/\Gamma$ where $\Gamma$ is the adiabatic index and $\gamma = (3\Gamma-1)/(\Gamma+1)$ is the effective adiabatic index in 2D.
The radial wavenumber of the most unstable mode is simply
\begin{equation}
    |k_r| = \frac{\pi G \Sigma}{c_{\rm s,eff}^2 + 2\pi G\Sigma H}.\label{eq:kr}
\end{equation}
Note that this is independent of $\omega$. In contrast, in a stable disk $k_r$ would depend on $\omega$ (which is set by the forcing that excites the wave).

Using Eq. \ref{eq:kr}, we predict the spiral pattern $\phi(r)=\int k_r{\rm d}r$ using our model's estimates of surface density and temperature profiles, and it shows good agreement with observation (Fig. \ref{fig:spiral} panel d).
Here we consider this analytic result (black lines in Fig. \ref{fig:spiral} panel d) as a prediction of our model because our axisymmetric model is blind to the phase of the spiral substructure. In other words, this agreement is not because we tuned our model to reproduce the observed pitch angle. This test suggests that the assumptions we take when translating observation to physical properties of the disk are likely appropriate (because otherwise this agreement would be merely coincidental, which is statistically unlikely).
Additionally, this serves as one of the strongest evidence of a gravitationally excited spiral substructure in protostellar/protoplanetary disks to date. While spiral substructures in a couple of systems have been attributed to GI due to qualitative evidences such as a relatively high estimated disk mass \citep[e.g.,][]{Lee2020,Veronesi2021} and/or morphological and kinematic signatures \citep[e.g.,][]{Forgan2018,Paneque2021},
this is the first time for a system to demonstrate \textit{quantitative agreement} between morphological signature (pitch angle) and independent estimates of surface density and temperature profiles.

\subsection{Spiral amplitude: why are spirals so hard to see?}\label{sec:spiral_amplitude}

Another important property of the spiral substructure is its amplitude. Although spirals in gravitationally self-regulated disks have been demonstrated in a number of simulations \citep[e.g.,][]{VorobyovBasu2007,XK21a,XK21b}, it remains challenging to make quantitative comparison between simulations and observations. The intrinsic variability of spirals \citep[e.g.,][Fig. 3]{XK21b} requires their properties (e.g., amplitude and radial extent) to be measured as statistical distributions; yet detailed, quantitative measurements of these distributions and how they depend on disk parameters (e.g., disk size, infall rate) are still absent. In the discussion below, we focus on characterizing the observational properties of the spiral in TMC1A; we also make a few qualitative comparisons with theory and simulations of spirals in gravitationally self-regulated disks.

Fig. \ref{fig:spiral} panel c shows that the amplitude of the spiral residual is $\sim 0.1$~K, which is $\lesssim 1\%$ of the peak of dust emission ($\sim 30~K$).
However, this does not mean that the underlying surface density perturbation has equally small relative amplitude.
In Fig. \ref{fig:spiral_amplitude} we compare the observed amplitude of the $m=1$ mode with the azimuthally averaged $T_{\rm b}$ profile. Both profiles are computed with the deprojected 0.9~mm image with circular synthesized beam in Fig. \ref{fig:spiral} panel b; the amplitude of the $m=1$ mode is obtained by a Fourier decomposition of the deprojected image along the azimuthal direction.

In the outer part of the disk ($\gtrsim 150$~au), the spiral has order-unity relative amplitude. The high relative amplitude of this $m=1$ spiral in the outer disk also causes the outer part of the disk to appear skewed (cf. 0.9~mm and 1.3~mm images in Fig. \ref{fig:obs_vs_model}). However, the absolute amplitude of the spiral remains low compared to the peak of emission because the outer part of the disk is very dim. This is mainly a result of the steep radial profiles of $\Sigma$ and $T_{\rm mid}$ (Section \ref{sec:results_overview}; see also Section \ref{sec:apparent}).

The observed relative amplitude of the spiral quickly decreases toward smaller radii. This is, however, largely due to observational factors which cause the observed $T_{\rm b}$ perturbation to have lower relative amplitude than the underlying surface density perturbation. One such observational factor is optical depth. When the emission is optically thick, only the surface of the disk is visible and the large-scale spirals can become almost invisible (see an example in \citealt{Xu22} Section 8.1). In Fig. \ref{fig:spiral_amplitude} we show the region with line-of-sight optical depth $\geq$1, 3, and 10 (grey shades); this affects the innermost $\sim 40$~au of the disk, explaining the low observed relative amplitude there. For this region, we cannot reliably constrain the amplitude of underlying column density fluctuation.
Another important observational factor is finite resolution. When the spiral is not well-resolved, the synthesized beam covers $\gtrsim 1$ wavelength and reduces the observed amplitude. The importance of this resolution effect can be estimated as follows. Locally, the spiral can be roughly approximated as a plane wave with wavenumber
\begin{equation}
k = \sqrt{k_r^2 + m^2/r^2}.
\end{equation}
Convolving this plane wave with a circular Gaussian beam with width $\sigma$ reduces the amplitude of wave by a factor of
\begin{align}
f_{\rm beam} &= \int_{-\infty}^{\infty} \frac{1}{\sigma\sqrt{2\pi}}\exp\left(-\frac{x^2}{2\sigma^2}\right)\cos(kx){\rm d}x\nonumber\\ &= \exp\left(-\frac 12 k^2\sigma^2\right).
\end{align}
Here $k$ can be computed from the $k_r$ in Eq. \ref{eq:kr}, and the $1\sigma$ width of the synthesized beam in the deprojected image (Fig. \ref{fig:spiral} panel b) is $25$~au (FWHM 59~au).
We can estimate the true amplitude of $T_{\rm b}$ perturbation by multiplying the observed amplitude by $1/f_{\rm beam}$ (red dashed lines in Fig. \ref{fig:spiral_amplitude}). In the optically thin region ($\gtrsim 60$~au), this reflects approximately the relative amplitude of the underlying surface density perturbation. 

After correcting for finite resolution, the relative amplitude of the spiral remains $\gtrsim 20\%$ for the optically thin region of the disk. We still see some radial variation of the spiral amplitude, which probably comes from the following two physical factors. First, in a gravitationally self-regulated disk, the typical relative amplitude of spiral waves decreases toward smaller radii (e.g., \citealt{VorobyovBasu2007} Fig. 2) which can be interpreted as a result of the radial profile of effective viscosity $\alpha_{\rm GI}$ (e.g., \citealt{XA23} Fig. 1). Additionally, the spiral waves in a graitationally self-regulated disk are generally not stationary and their amplitudes can exhibit significant variability. The dip of spiral amplitude at $\sim 120$~au might be associated with such stochastic fluctuation.

In summary, the low observed amplitude of spiral in TMC1A can be associated with a number of physical and observational factors, and the underlying surface density perturbation may in fact have high relative amplitude (cf. red dashed line in Fig. \ref{fig:spiral_amplitude} bottom panel). It is also worth noting that even when the spiral is resolved and has high relative amplitude (as in the outer part of TMC1A), it could be hard to determine visually whether the image contains a spiral or some generic skew due to other reasons (e.g., envelope contamination) unless one deprojects the image and inspect the non-axisymmetric residual. This might explain the lack of easily visible spirals in existing survey of Class 0/I disks \citep[e.g.,][]{Tobin2020}.

\section{Grain size: modest grain growth in line with fragmentation barrier}\label{sec:dust}

\begin{figure}
    \centering
    \includegraphics[scale=0.6]{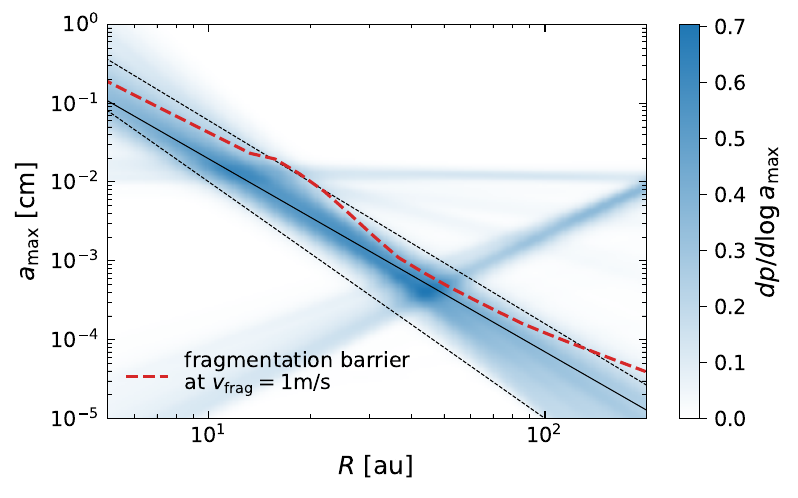}
    \caption{Posterior distribution of the grain size ($a_{\rm max}$). Profiles corresponding to fiducial values and 68\% confidence intervals of $a_{\rm max}^{10{\rm au}}$ and $a_{\rm max}^{100{\rm au}}$ are shown in solid and dashed black lines, respectively. The fragmentation barrier for our disk model at fragmentation threshold velocity $v_{\rm frag}=1{\rm m/s}$ is plotted for reference (red dashed line).
    The grain size remains small beyond a few au, and the main branch of the posterior shows steep radial dependence consistent with the fragmentation barrier.
    }
    \label{fig:grain_size}
\end{figure}

\begin{figure*}
    \centering
    \includegraphics[scale=0.6]{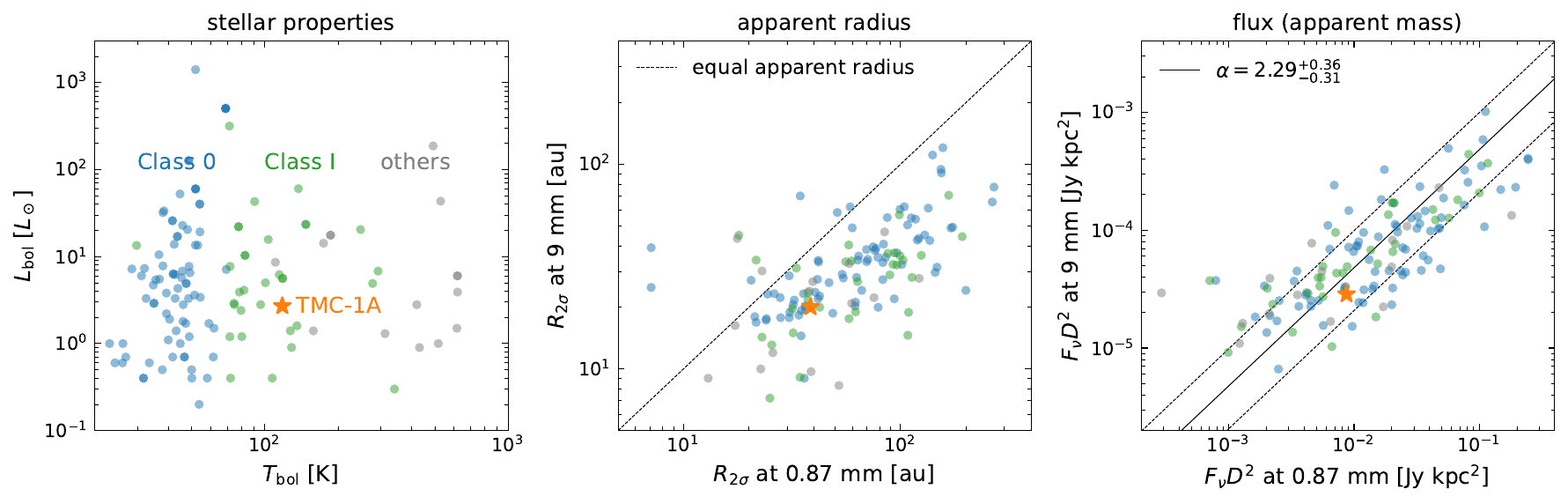}
    \caption{Comparison between observational properties of TMC1A (this work and \citealt{Kristensen2012}) and the disk population from the VANDAM Orion survey \citep{Tobin2020}. To make an apple-to-apple comparison, we follow the definitions in \citet{Tobin2020} and compute apparent radius and flux from the best-fit Gaussian profile. (Apparent radius is defined as $2\sigma$ of the Gaussian.)
    For the last panel, we label the median and 16/84th percentiles of the spectral index distribution of the VANDAM Orion sample for reference. All properties of TMC1A are quite typical compared to those of the VANDAM Orion sample.}
    \label{fig:population}
\end{figure*}

Fig. \ref{fig:grain_size} shows the $a_{\rm max}$ posterior of our model. The distribution shows some degeneracy and contains a few branches, but all these branches show small ($<$mm) $a_{\rm max}$ beyond a few au. The fiducial values of $a_{\rm max}^{10\rm au}$ and $a_{\rm max}^{100\rm au}$ correspond to a radial profile of (black solid line in Fig. \ref{fig:grain_size})
\begin{equation}
    a_{\rm max} \sim 196(R/10\rm au)^{-2.45} \mu{\rm m}\label{eq:amax}.
\end{equation}
Relaxing the assumption of marginal instability, as discussed in Section \ref{sec:test}, barely affects these results.

Our results allow an independent comparison between grain growth theory and observation. Note that our model makes no assumption on how disk properties affect grain size.
Theoretically, collisional growth of dust grains yield a maximum grain size similar to the fragmentation barrier, where typical collisions between dust grains lead to fragmentation \citep{Birnstiel2010}.
Using the radial profile of the surface density, temperature, and turbulent $\alpha$ (which can be estimated from the accretion rate), we compute the fragmentation barrier in Fig. \ref{fig:grain_size} (see details of the calculation in Appendix \ref{a:fragmentation}).
The main branch of the $a_{\rm max}$ posterior shows remarkable agreement with a fragmentation barrier at fragmentation threshold velocity $v_{\rm frag}\sim 1~{\rm m/s}$.
The fragmentation barrier naturally explains the large variation of $a_{\rm max}$ across the disk: in a gravitationally self-regulated disk, $\alpha_{\rm GI}$ tends to be radially increasing (cf. Fig. 1 in \citealt{XA23}), making the radial dependence of $a_{\rm max}$ steeper compared to models that assume constant turbulent $\alpha$ (e.g., \citealt{Birnstiel2010}).

Previously, \citet{Aso2021} find that the dust continuum polarization in the central region of the TMC1A disk shows morphological features resembling optically thick dust self-scattering. The peak polarization fraction is $\sim 1\%$, which suggests intermediate-sized grains with $a_{\rm max} = 80$--$300~\mu$m. In our model, this $a_{\rm max}$ range corresponds to an annulus with inner radius 7--13~au and outer radius 11--22~au. (These ranges correspond to the 68\% confidence interval of our grain size estimate, which is shown in black dotted lines in Fig. \ref{fig:grain_size}.) This is broadly consistent with the observed size of the central polarization region, which extends to a radius of $\sim 50$~au at $\sim 50$~au synthesized beam width. Future observations with higher resolution would allow a more detailed comparison.
Meanwhile, using the absence of molecular line emission, \citet{Harsono2018} concluded that TMC1A either hosts $\gtrsim$mm grains or is massive and gravitationally unstable, although they favored the former explanation because at that time there was no strong evidence for GI (the spiral in Fig. \ref{fig:spiral} had not been detected).
The qualitative trend that grain size quickly increases towards small radii is also consistent with observational evidences of $\gtrsim$mm grains in the innermost $\sim10$ au of FU Ori \citep{Liu2021b}, whose outburst behavior may be related to GI \citep{VorobyovBasu2010,Zhu2010}.
On the other hand, we find much smaller grains than the predictions from recent 2D simulations of grain growth in gravitationally self-regulated disks \citep{Vorobyov2019,Elbakyan2020,Vorobyov2023}, and that discrepancy is in part because they assume a much higher $v_{\rm frag}$ ($3\sim 30~{\rm m/s}$). Additionally, in these studies the turbulent $\alpha$, which affects the fragmentation barrier, is modeled as a free (and often small) parameter independent of the GI dynamics.
Meanwhile, our estimate of the fragmentation barrier assumes $\alpha\sim\alpha_{\rm GI}$ on the ground that 3D simulations (e.g., \citealt{RiolsLatter2017}) find that a nontrivial fraction of dissipation in a gravito-turbulence occurs through a turbulent cascade.

\begin{figure}
    \centering
    \includegraphics[scale=0.6]{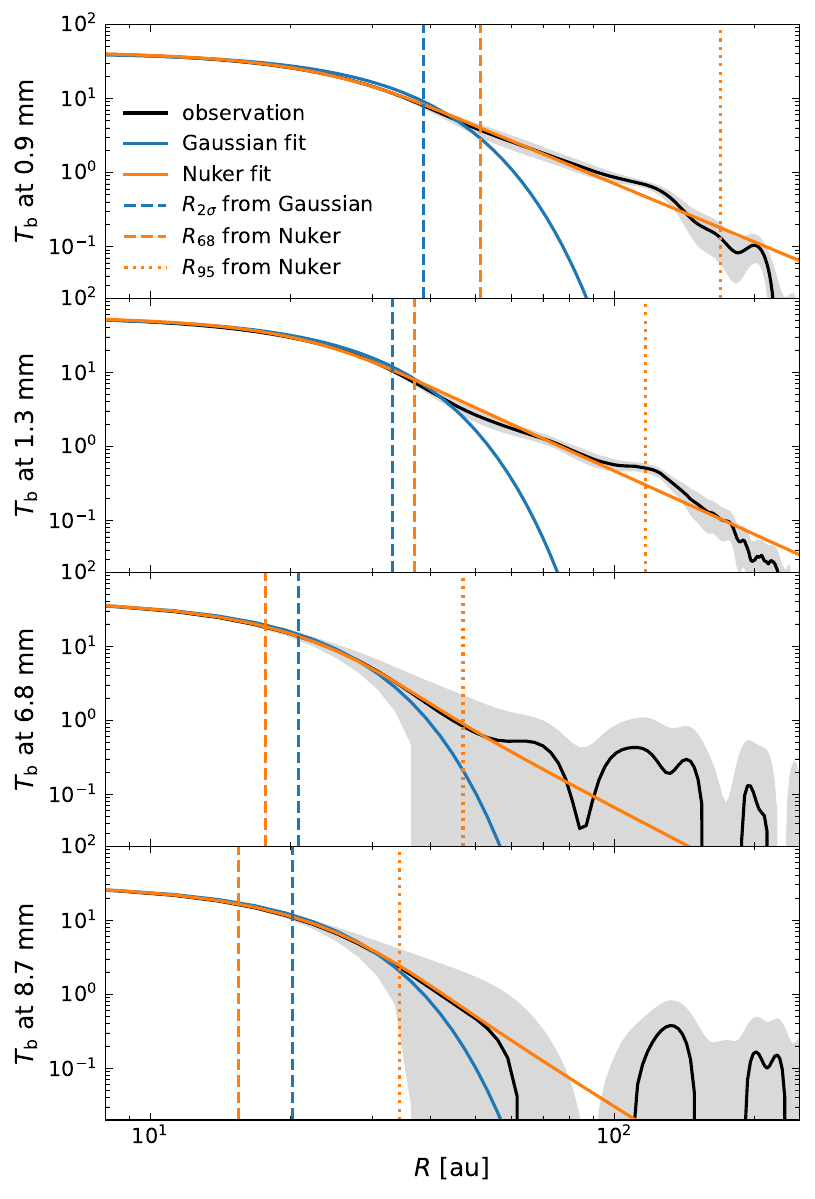}
    \caption{Apparent disk size of TMC1A at different wavelengths. We show three commonly used definitions of apparent size: $2\sigma$ from a Gaussian fit and 68 and 95th percentiles from a Nuker fit. Because the emission profile is steep in the optically thin outer disk, the apparent disk size is generally smaller than the actual disk size, and shows a trend to decrease towards longer wavelengths.}
    \label{fig:apparent_size}
\end{figure}

\section{Discussion}\label{sec:discussion}
\subsection{Is TMC1A special?}\label{sec:typical}

The observations and our modeling of TMC1A show a large, massive disk with a prominent gravitationally excited spiral arm. Meanwhile, observational estimates from surveys of Class 0/I disks \citep[e.g.,][]{Tobin2020,Sheehan2022} suggest that these disks are generally small (estimated size $\sim 40$au), low-mass (estimated disk mass $\sim 0.01M_\odot$), and do not often show substructure. That seems to suggest that the disk around TMC1A is just a special case unrepresentative of the general population of Class 0/I disks.

However, this is not the case. In Fig. \ref{fig:population} we do an apple-to-apple comparison between the disk around TMC1A and the Class 0/I disks from the VANDAM Orion survey \citep{Tobin2020}. For key observational diagnostics including bolometric luminosity and temperature, apparent disk size, flux (often used to give apparent mass estimates), and spectral index, TMC1A shows values similar to those of other sources in the survey. In other words, judging by known observational metrics the protostellar disk around TMC1A is a fairly typical Class 0/I disk. The difference in estimated disk properties likely originates from different assumptions for modeling the observation (see next subsection).

\subsection{Implications on interpreting disk observations}\label{sec:apparent}

Our model of TMC1A shows a relatively steep density profile (Fig. \ref{fig:model_summary}). This is a generic (but not necessarily unique) feature of gravitationally self-regulated disks, which shows approximately $\Sigma\propto R^{-2}$ \citep{Xu22}, and it leads to a few interesting observational properties.

The radial variation of surface density (and temperature) means that the disk is generally optically thick (to observation) at small radii and optically thin at sufficiently large radii. For a steep surface density profile, most disk mass is in the inner part of the disk, which remains optically thick. As a result, only a small fraction of mass is visible, and the apparent mass of the disk can be much lower than the actual mass. Another implication of the steep surface density profile is that the radial profile of emission is very steep in the optically thin outer disk, thus the outer region of the disk makes little contribution to the total flux (Fig. \ref{fig:apparent_size}). As a result, the apparent disk size -- often estimated using the contour containing a certain fraction of the total flux -- could be much smaller than the actual disk size (i.e., the transition radius between infall and Keplerian rotation).

Additionally, for disks with such steep radial profiles the visible disk mass and apparent disk size should both scale with the opacity of the tracer (at the observed wavelength). For more optically thin tracers, a larger portion of the disk mass becomes visible. This naturally explains why dust continuum emission shows low spectral index \citep[cf.][]{Xu22}. Low spectral index is traditionally interpreted as low dust opacity index and therefore large grain size, but this interpretation is in tension with constraints from polarization (see a discussion in \citealt{Liu2021b}). This mechanism also reconciles the tension between apparent mass obtained from different tracers, including dust continuum, CO, HD \citep[e.g.,][]{McClure2016}, and rare isotopes such as $^{13}$C$^{17}$O \citep[e.g.,][]{Booth2019}. Such tension is traditionally interpreted as under/overestimated tracer abundances. 

A similar trend exists for apparent disk size. For more optically thin tracers, a larger range of radii becomes optically thin and show steep emission profile, causing the apparent disk size to decrease as shown in Fig. \ref{fig:apparent_size}. This naturally explains why the apparent size decreases from shorter to longer wavelengths \citep[e.g.,][]{Tobin2020} and from CO to dust continuum \citep[e.g.,][]{Ansdell2018}.
Such trend is often interpreted as a result of radial dust drift or concentration of dust grains into unresolved, optically thick clumps with volume filling factor of a few tens of percent (\citealt{Tripathi2018,Tazzari2021tension}; also see \citealt{Tripathi2017,Andrews2018}).

In summary, using TMC1A as an example (cf. Fig. \ref{fig:population} and \ref{fig:apparent_size}) we highlight the possibility that many observed trends traditionally interpreted as low disk mass, small disk size, large dust grains, and under/overestimated tracer abundances may instead be (mainly or partly) due to a steep radial disk profile. We plan to investigate this possibility more quantitatively in future studies.

\subsection{Implications on the onset of planet formation}
If typical Class 0/I disks have high mass and small grains like TMC1A, it would be an important constraint for when and how the first stages of planet formation occur. In particular, given that we seldom (if ever) reliably detect large grains in young disks, the first stage of planet formation may not involve significant increase in \textit{typical} grain size. Instead, it may be facilitated by the formation of a bimodal grain size distribution \citep{Windmark2012, XA23} and/or direct gravitational collapse into planetesimals in dust clumps formed by instabilities and turbulence (e.g., streaming instability \citealt{YoudinGoodman2005}; also see reviews in \citealt{Klahr2018,Drazkowska2022}). In that case, the grains (and planetesimals) above the fragmentation barrier may only contain a small fraction of total dust mass (and barely contribute to dust continuum emission) but, given the high disk mass, could dominate the planet (solid) mass budget \citep[e.g.,][Fig. 10]{XA23}.

\section{Conclusion}\label{sec:conclusion}
In this paper we model multi-wavelength dust continuum observations of the disk around the Class I protostar TMC1A
to constrain the disk's surface density and temperature profile and the level of dust growth. Dust continuum observations of the disk can be fit well by a gravitationally self-regulated model which is marginally gravitationally unstable everywhere and is mainly heated by the effective viscosity of GI (Section \ref{sec:model} and \ref{sec:results}).
The surface density and midplane temperature profiles produced by our model are summarized in Fig. \ref{fig:model_summary} and Eqs. \ref{eq:Sigma}--\ref{eq:Tmid}. TMC1A contains a $m=1$ spiral substructure, and the pitch angle predicted from our model agrees with the observed spiral geometry;
this serves as both a test of our model and strong evidence of the gravitational origin of the spiral (Section \ref{sec:spiral_pitch}).
The observed amplitude of the spiral substructure remains low (a few percent of the peak intensity), but the underlying surface density fluctuation can have large relative amplitude ($\gtrsim 20\%$); the low observed amplitude is the combined effect of the high optical depth at small radii, the low luminosity at large radii, and limited resolution (Section \ref{sec:spiral_amplitude}).
Our model also constrains the radial profile of maximum grain size $a_{\rm max}$ (Fig. \ref{fig:grain_size} and Eq. \ref{eq:amax}); our estimated $a_{\rm max}$ profile is consistent with fragmentation barrier at fragmentation threshold $\sim 1{\rm m}/{\rm s}$ (Section \ref{sec:dust}).
Our modeling of TMC1A has broader implications on the interpretation of existing and future observations of young disk populations (Section \ref{sec:discussion}). We demonstrate that the disk around TMC1A is a very typical Class 0/I disks in terms of observational metrics.
Moreover, we show that the low flux, low spectral index, and small and wavelength-dependent apparent size observed ubiquitously in young disk populations (including TMC1A), which are traditionally interpreted as signatures of small disk size, low dust mass, and large dust grains, could instead be produced by the radius- and wavelength-dependence of optical depth alone.
This suggests a possibility that existing interpretations of dust continuum observations might systematically underestimate disk mass and size and overestimate dust grain size, which would have important implications for planet formation.

~

This paper makes use of the following ALMA data: ADS/JAO.ALMA\#2015.1.01415.S, 
ADS/JAO.ALMA\#2015.1.01549.S, 
and\\ 
ADS/JAO.ALMA\#2018.1.00701.S. ALMA is a partnership of ESO (representing its member states), NSF (USA) and NINS (Japan), together with NRC (Canada), MOST and ASIAA (Taiwan), and KASI (Republic of Korea), in cooperation with the Republic of Chile. The Joint ALMA Observatory is operated by ESO, AUI/NRAO and NAOJ.
The National Radio Astronomy Observatory is a facility of the National Science Foundation operated under cooperative agreement by Associated Universities, Inc.
H.B.L. is supported by the National Science and Technology Council (NSTC) of Taiwan (Grant Nos. 111-2112-M-110-022-MY3).
S.O. is supported by a Grant-in-Aid from Japan Society for the Promotion of Science (KAKENHI: Nos. JP20K14533, JP20H00182, JP22H01275).
W.X. thanks Cassandra Hall, Jiayin Dong, Mordecai-Mark Mac Low, and Philip Armitage for insightful discussions.

\software{
\texttt{emcee} \citep{Foreman-Mackey2013},
\texttt{RADMC-3D} \citep{Dullemond2012}
}

\appendix
\section{Details of the MCMC fitting}\label{a:fit}

In this appendix we discuss the priors and the log-likelihood function used for our MCMC fit. 

We choose broad, uninformative priors for all variables (except $Q$ and $q$) to avoid having our estimates affected by the priors. For disk orientation, we choose uniform priors with $\cos i \in [0,1]$, $\theta\in [0,\pi]$. For other physical disk parameters, we choose log-uniform priors with $M_{\rm tot}\in [0.1,10]~{\rm M}_\odot$, $R_{\rm d}\in [20,300]~{\rm au}$, $\dot M\in[10^{-7},10^{-4}]~{\rm M}_\odot/{\rm yr}$, 
$a_{\rm max}^{10{\rm au},100{\rm au}} \in [10^{-5}, 1]~{\rm cm}$, $Q^{10{\rm au},100{\rm au}}\in[1,2]$, and $q\in [2.5,3.5]$.
Adopting the assumption of marginal instability, we further trim the prior distribution to only consider parameters satisfying $Q\in[1,2]$ everywhere between 5 au and $R_{\rm d}$. We also fit another model without this assumption as described in Section \ref{sec:test}.

The log-likelihood function $l$ of a given set of parameters contains contributions from all four wavelengths as well as the dynamical mass constraint, and is given by (up to a constant offset which does not affect the result)
\begin{equation}
    l = l_{\rm p} + \sum_{i=1}^{4} l_{\lambda_i}.
\end{equation}
Here the dynamical mass constraint corresponds to
\begin{equation}
    l_{\rm p} = -\frac{[M_{\rm tot}\sin^2 i - M_{\rm p}]^2}{2\sigma_{M_{\rm p}}^2},
\end{equation}
with $M_{\rm p} = 0.56~{\rm M}_\odot$ and $\sigma_{M_{\rm p}} = 0.05~{\rm M}_\odot$ (Eq. \ref{eq:Mp}).
The constraints from dust continuum correspond to (similar to Eqs. 9--10 in \citealt{Xu22})
\begin{equation}
    l_{\lambda_i} = -\int \left[\frac{|T_{{\rm b, obs},i}-T_{{\rm b,model},i}|^2}{2\sigma^2} + \frac 12 \log(\sigma^2)\right] S_{B,i}^{-1} {\rm d}S,\label{eq:li}
\end{equation}
with the uncertainty given by
\begin{equation}
    \sigma^2 = \sigma_{{\rm obs},i}^2+\sigma^2_{\rm rel,~model}T_{{\rm b,model},i}^2.
\end{equation}
Here $T_{\rm b,obs}, T_{\rm b,model}$ are brightness temperatures from observations and from model prediction (after convolving with the synthesized beam), $S_{\rm B}$ is the synthesized beam size of the observation, $\sigma_{\rm obs}$ is the observational uncertainty in $T_{\rm b}$, and $\sigma_{\rm rel, model}$ is the relative uncertainty of the model.
We discuss how we obtain $\sigma_{\rm obs,i}$ and $\sigma_{\rm rel, model}$ in the next few paragraphs. The integration goes over the whole image, with ${\rm d}S$ and $S_{\rm B}$ both in unit of physical area (e.g., au$^2$). Eq. \ref{eq:li} can be interpreted as a summation of the log likelihood of getting $T_{\rm b,obs}$ over each synthesized beam; the term in the brackets corresponds to the logarithm of a Gaussian distribution of $T_{{\rm b, obs},i}-T_{{\rm b,model},i}$ with variance $\sigma^2$.

$\sigma_{\rm rel, model}$ intends to capture the error due to oversimplifications in our model assumption.
More precisely, we define $\sigma_{\rm obs}$ as errors due to asymmetric perturbations in the disk (which our axiysmmetric model cannot capture) and observational noise, whereas $\sigma_{\rm rel, model}$ captures the errors in the axisymmetric profile of the disk. Such errors could originate from, for instance, the deviation of the radial profiles of $Q$ and $a_{\rm max}$ from exact power-laws and radial variation of $\dot M$ and $q$.
Because of our limited understanding on disk evolution, it would be difficult to estimate accurately the amplitude of this error a priori. In particular, while one can attempt to estimate this error by comparing against a simulation, it is often unclear how much error the simulation contains due to its limited physical ingredients and numerical resolution; such a comparison would be more of a sanity check than a reliable estimate of error.
Given this limitation, we choose to treat $\sigma_{\rm rel, model}$ as an uncertainty of unknown amplitude.
In Bayesian analysis, such uncertainty can be treated as a ``nuisance parameter'', which is included as a free parameter in the model and marginalized over (see a tutorial by \citealt{Hogg2010}).
This also helps determining whether the model is consistent with the data; if the $\sigma_{\rm rel, model}$ posterior favors values much larger than the theoretically expected inaccuracy of the model, it suggests that some assumptions of the model may be inappropriate.
We choose a uniform prior of $\sigma_{\rm rel, model}\in[0,1]$, which is much broader than the posterior distribution.
In our fiducial model the posterior of $\sigma_{\rm rel, model}$ peaks at $\sim 0.2$;
this is broadly consistent with the level of difference when we compare (a slightly simpler version of) our model to a 3D disk simulation in \citet{XK21b}.

Choosing $\sigma_{\rm obs}$ is more tricky. An intuitive choice would be to follow \citet{Xu22} and use the RMS error of the observation, $\sigma_{\rm rms}$, as $\sigma_{\rm obs}$.
This can already produce reasonable estimates for the model parameters, but we caution that it could overestimate the disagreement between model and observation $(-l_{\lambda_i})$, which would then result in a problematic underestimation of parameter uncertainty (posterior width) in our Bayesian fit.
The potential overestimation of $-l_{\lambda_i}$ comes from two issues.
First, $\sigma_{\rm rms}$ often underestimate the amplitude of errors introduced during image reconstruction in brighter regions (in this case the disk). 
Second, as we mentioned previously, $\sigma_{\rm obs}$ is intented to capture the expected difference between the noiseless, azimuthally averaged disk profile and the actual observation. Since we do resolve the spiral in TMC1A at shorter wavelengths, its amplitude should be incorporated in $\sigma_{\rm obs}$ in addition to $\sigma_{\rm rms}$.
In order to account for these issues, we define $\sigma_{\rm obs}$ as follows.
We first deproject the disk and measure the standard deviation of the deprojected $T_{\rm b}$ in the azimuthal direction at a given radius, $\sigma_{\rm asym}$. The we define
\begin{equation}
    \sigma_{\rm obs}\equiv \max\{\sigma_{\rm asym}, \sigma_{\rm rms}\}.
\end{equation}
This definition naturally includes both the random fluctuations due to observational noise and image reconstruction and any asymmetric perturbation (mainly the spiral) that cannot be captured by our axiysmmetric model.

We comment that it is crucial to have accurate estimate of (or leave sufficient freedom for) various components of the uncertainty $\sigma^2$, especially in a Bayesian model.
In particular, when we underestimate $\sigma^2$ -- which might be common in the literature since the uncertainty of the model is often ignored -- it makes the log likelihood drop faster around the best-fit parameters, causing the delusion that the model gives better constraints on the parameters than it actually can. Ignoring the model uncertainty could also bias the model by overfitting the part of the observation that has the least observational uncertainty, while in reality the total uncertainty may be dominated by the model uncertainty and more uniform.

\section{Radiative transfer modeling}\label{a:radmc}
The radiative transfer calculations shown in the bottom row of Fig. \ref{fig:obs_vs_model} and Fig. \ref{fig:sup_radmc} are performed using \texttt{RADMC-3D} \citep{Dullemond2012}, with disk properties based on our fiducial disk model. The \texttt{RADMC-3D} calculation takes a 3D density profile and relevant heating and radiation sources to self-consistently compute equilibrium temperature and emission. The density profile is given by our model's surface density profile together with a simple vertical density profile of $\rho(z) \propto \exp(-z^2/2H^2)$ where $H=c_s/\Omega$ is the disk scale height (with $c_s$ computed from our model's vertically averaged temperature). The heating sources consist of internal gravito-viscous heating due to GI and external heating due to protostellar irradiation. For the former, where we adopt the prescription in \citet{Xu22} and assume heating per unit radius is $\sim -g_R \dot M$ where $g_R$ is the radial gravity. In general this heating rate contains an order-unity prefactor. For viscous heating (or, more generally, any local transport in a thin disk) this prefactor is $3/2$; but for transport and heating by gravitational instability, the finite disk thickness and the (semi-)global nature of spiral waves preclude a precise estimate (see Appendix D of \citealt{XK21b}). For the latter, we adopt an effective temperature of $T_{\rm eff}=4300~{\rm K}$ and choose the stellar radius to match the observed bolometric luminosity of $2.7~{\rm L}_\odot$ \citep{Kristensen2012}.

\section{The fragmentation barrier}\label{a:fragmentation}

The fragmentation barrier shown in Fig. \ref{fig:grain_size} has been computed as follows.
Our disk model produces the radial profile of the temperature (sound speed) and surface density. These, together with the accretion rate, can be used for estimating the effective viscosity $\alpha_{\rm GI}$ which characterizes the level of turbulence via
\begin{equation}
    \alpha_{\rm GI} = \frac{\dot M\Omega}{3\pi\Sigma c_s^2}.
\end{equation}
Note that estimating $\alpha_{\rm GI}$ with the level of turbulent heating would give a similar result, since our model implicitly assumes that the effective viscosity for heating and angular-momentum transport (accretion) are similar. (In reality, they are not necessarily the same but should be comparable; see discussion in \citealt{XK21b}.)
The typical relative velocity between dust grains are set by the grain size and $\alpha_{\rm GI}$ (Eqs. 28--29 in \citealt{OrmelCuzzi2007})
\begin{equation}
    v_{\rm t}^2 \sim \min\left\{\frac 92 \alpha_{\rm GI} c_s^2 {\rm St}, \frac 32 \alpha_{\rm GI} c_s^2 \left(1+\frac{1}{1+{\rm St}}\right)\right\}.\label{eq:vt}
\end{equation}
The Stokes number St is related to the grain size $a$ by (Eq. 2 in \citealt{Birnstiel2012})
\begin{equation}
    {\rm St} = \frac{a\rho_{\rm s}}{\Sigma}\frac{\pi}{2}.
\end{equation}
Here $\rho_{\rm s}$ is the density of the grain, which we compute following \citet{dsharpV}, in consistency with our dust opacity model.
The grain size at fragmentation barrier can be computed by setting $v_{\rm t}$ equal to the threshold velocity of collisional fragmentation, $v_{\rm frag}$, which we take to be $1~{\rm m/s}$.

There are some uncertainties regarding the exact value of $v_{\rm frag}$, and it is not necessarily constant throughout the disk. For our modeling, the choice of $1~{\rm m/s}$ follows the convention of grain coagulation calculations (e.g., \citealt{Birnstiel2010}) and is broadly consistent with grain collision experiments (see a review in \citealt{BlumWurm2008}).
One caveat is that the estimate of fragmentation threshold shows large variation across different studies.
Two extreme examples are $\sim 0.2~{\rm m/s}$ from silicate collision experiments (e.g., \citealt{Beitz2011}) and several $10~{\rm m/s}$ from ice collision simulations (e.g., \citealt{Wada2009}), although more recent results by \citet{Gundlach2018} seem to suggest an underestimation in the former case (due to the choice of large monomer size) and an overestimation in the latter (due to a possible overestimate of tensile strength); this is also consistent with the result from \citet{Musiolik2019} that the material property of grains with and without water ice may, after all, be quite similar (with the potential exception of a narrow region around the snowline).
Another potentially important effect is that ``drying'' silicates (i.e., removing the surface water), which happens at $\gtrsim500$~K (cf. \citealt{DAngelo2019}), could significantly increase the fragmentation barrier \citep{Kimura2015,Steinpilz2019,Pillich2021}. However, the current observation would be relatively insensitive to further increase of grain size at the small radii required for such hot temperature.

In the current study our focus is to show that our grain size constraint is (broadly) compatible with existing theory and experiment of collisional grain growth.
In the future, one might instead use observational constraints of grain size to place a tight constraint on $v_{\rm frag}$ (cf. \citealt{Yamamuro2023}), at least in the resolved outer disk.
This seems especially promising in gravitationally self-regulated disks as the strong physical constraints of gravitational self-regulation reduces the uncertainties in disk properties, especially the effective $\alpha$ of turbulence.


\bibliography{bib}{}

\begin{thebibliography}{}
\expandafter\ifx\csname natexlab\endcsname\relax\def\natexlab#1{#1}\fi
\providecommand{\url}[1]{\href{#1}{#1}}
\providecommand{\dodoi}[1]{doi:~\href{http://doi.org/#1}{\nolinkurl{#1}}}
\providecommand{\doeprint}[1]{\href{http://ascl.net/#1}{\nolinkurl{http://ascl.net/#1}}}
\providecommand{\doarXiv}[1]{\href{https://arxiv.org/abs/#1}{\nolinkurl{https://arxiv.org/abs/#1}}}

\bibitem[{{Andrews}(2020)}]{Andrews2020}
{Andrews}, S.~M. 2020, \araa, 58, 483,
  \dodoi{10.1146/annurev-astro-031220-010302}

\bibitem[{{Andrews} {et~al.}(2018{\natexlab{a}}){Andrews}, {Terrell},
  {Tripathi}, {Ansdell}, {Williams}, \& {Wilner}}]{Andrews2018}
{Andrews}, S.~M., {Terrell}, M., {Tripathi}, A., {et~al.} 2018{\natexlab{a}},
  \apj, 865, 157, \dodoi{10.3847/1538-4357/aadd9f}

\bibitem[{{Andrews} {et~al.}(2018{\natexlab{b}}){Andrews}, {Huang},
  {P{\'e}rez}, {Isella}, {Dullemond}, {Kurtovic}, {Guzm{\'a}n}, {Carpenter},
  {Wilner}, {Zhang}, {Zhu}, {Birnstiel}, {Bai}, {Benisty}, {Hughes},
  {{\"O}berg}, \& {Ricci}}]{dsharpI}
{Andrews}, S.~M., {Huang}, J., {P{\'e}rez}, L.~M., {et~al.} 2018{\natexlab{b}},
  \apjl, 869, L41, \dodoi{10.3847/2041-8213/aaf741}

\bibitem[{{Ansdell} {et~al.}(2018){Ansdell}, {Williams}, {Trapman}, {van
  Terwisga}, {Facchini}, {Manara}, {van der Marel}, {Miotello}, {Tazzari},
  {Hogerheijde}, {Guidi}, {Testi}, \& {van Dishoeck}}]{Ansdell2018}
{Ansdell}, M., {Williams}, J.~P., {Trapman}, L., {et~al.} 2018, \apj, 859, 21,
  \dodoi{10.3847/1538-4357/aab890}

\bibitem[{{Aso} {et~al.}(2021){Aso}, {Kwon}, {Hirano}, {Ching}, {Lai}, {Li}, \&
  {Rao}}]{Aso2021}
{Aso}, Y., {Kwon}, W., {Hirano}, N., {et~al.} 2021, \apj, 920, 71,
  \dodoi{10.3847/1538-4357/ac15f3}

\bibitem[{{Aso} {et~al.}(2015){Aso}, {Ohashi}, {Saigo}, {Koyamatsu}, {Aikawa},
  {Hayashi}, {Machida}, {Saito}, {Takakuwa}, {Tomida}, {Tomisaka}, \&
  {Yen}}]{Aso2015}
{Aso}, Y., {Ohashi}, N., {Saigo}, K., {et~al.} 2015, \apj, 812, 27,
  \dodoi{10.1088/0004-637X/812/1/27}

\bibitem[{{Beitz} {et~al.}(2011){Beitz}, {G{\"u}ttler}, {Blum}, {Meisner},
  {Teiser}, \& {Wurm}}]{Beitz2011}
{Beitz}, E., {G{\"u}ttler}, C., {Blum}, J., {et~al.} 2011, \apj, 736, 34,
  \dodoi{10.1088/0004-637X/736/1/34}

\bibitem[{{Bertin} \& {Lodato}(1999)}]{BertinLodato1999}
{Bertin}, G., \& {Lodato}, G. 1999, \aap, 350, 694,
  \dodoi{10.48550/arXiv.astro-ph/9908095}

\bibitem[{{Birnstiel} {et~al.}(2010){Birnstiel}, {Dullemond}, \&
  {Brauer}}]{Birnstiel2010}
{Birnstiel}, T., {Dullemond}, C.~P., \& {Brauer}, F. 2010, \aap, 513, A79,
  \dodoi{10.1051/0004-6361/200913731}

\bibitem[{{Birnstiel} {et~al.}(2012){Birnstiel}, {Klahr}, \&
  {Ercolano}}]{Birnstiel2012}
{Birnstiel}, T., {Klahr}, H., \& {Ercolano}, B. 2012, \aap, 539, A148,
  \dodoi{10.1051/0004-6361/201118136}

\bibitem[{{Birnstiel} {et~al.}(2018){Birnstiel}, {Dullemond}, {Zhu}, {Andrews},
  {Bai}, {Wilner}, {Carpenter}, {Huang}, {Isella}, {Benisty}, {P{\'e}rez}, \&
  {Zhang}}]{dsharpV}
{Birnstiel}, T., {Dullemond}, C.~P., {Zhu}, Z., {et~al.} 2018, \apjl, 869, L45,
  \dodoi{10.3847/2041-8213/aaf743}

\bibitem[{{Blum} \& {Wurm}(2008)}]{BlumWurm2008}
{Blum}, J., \& {Wurm}, G. 2008, \araa, 46, 21,
  \dodoi{10.1146/annurev.astro.46.060407.145152}

\bibitem[{{Booth} {et~al.}(2019){Booth}, {Walsh}, {Ilee}, {Notsu}, {Qi},
  {Nomura}, \& {Akiyama}}]{Booth2019}
{Booth}, A.~S., {Walsh}, C., {Ilee}, J.~D., {et~al.} 2019, \apjl, 882, L31,
  \dodoi{10.3847/2041-8213/ab3645}

\bibitem[{{Chiang} \& {Goldreich}(1997)}]{ChiangGoldreich1997}
{Chiang}, E.~I., \& {Goldreich}, P. 1997, \apj, 490, 368,
  \dodoi{10.1086/304869}

\bibitem[{{Cossins} {et~al.}(2009){Cossins}, {Lodato}, \&
  {Clarke}}]{Cossins2009}
{Cossins}, P., {Lodato}, G., \& {Clarke}, C.~J. 2009, \mnras, 393, 1157,
  \dodoi{10.1111/j.1365-2966.2008.14275.x}

\bibitem[{{D'Angelo} {et~al.}(2019){D'Angelo}, {Cazaux}, {Kamp}, {Thi}, \&
  {Woitke}}]{DAngelo2019}
{D'Angelo}, M., {Cazaux}, S., {Kamp}, I., {Thi}, W.~F., \& {Woitke}, P. 2019,
  \aap, 622, A208, \dodoi{10.1051/0004-6361/201833715}

\bibitem[{{Draine}(2006)}]{Draine2006}
{Draine}, B.~T. 2006, \apj, 636, 1114, \dodoi{10.1086/498130}

\bibitem[{{Drazkowska} {et~al.}(2022){Drazkowska}, {Bitsch}, {Lambrechts},
  {Mulders}, {Harsono}, {Vazan}, {Liu}, {Ormel}, {Kretke}, \&
  {Morbidelli}}]{Drazkowska2022}
{Drazkowska}, J., {Bitsch}, B., {Lambrechts}, M., {et~al.} 2022, arXiv
  e-prints, arXiv:2203.09759.
\newblock \doarXiv{2203.09759}

\bibitem[{{Dullemond} {et~al.}(2012){Dullemond}, {Juhasz}, {Pohl}, {Sereshti},
  {Shetty}, {Peters}, {Commercon}, \& {Flock}}]{Dullemond2012}
{Dullemond}, C.~P., {Juhasz}, A., {Pohl}, A., {et~al.} 2012, {RADMC-3D: A
  multi-purpose radiative transfer tool}, Astrophysics Source Code Library,
  record ascl:1202.015.
\newblock \doeprint{1202.015}

\bibitem[{{Elbakyan} {et~al.}(2020){Elbakyan}, {Johansen}, {Lambrechts},
  {Akimkin}, \& {Vorobyov}}]{Elbakyan2020}
{Elbakyan}, V.~G., {Johansen}, A., {Lambrechts}, M., {Akimkin}, V., \&
  {Vorobyov}, E.~I. 2020, \aap, 637, A5, \dodoi{10.1051/0004-6361/201937198}

\bibitem[{{Foreman-Mackey} {et~al.}(2013){Foreman-Mackey}, {Hogg}, {Lang}, \&
  {Goodman}}]{Foreman-Mackey2013}
{Foreman-Mackey}, D., {Hogg}, D.~W., {Lang}, D., \& {Goodman}, J. 2013, \pasp,
  125, 306, \dodoi{10.1086/670067}

\bibitem[{{Forgan} {et~al.}(2018){Forgan}, {Ilee}, \& {Meru}}]{Forgan2018}
{Forgan}, D.~H., {Ilee}, J.~D., \& {Meru}, F. 2018, \apjl, 860, L5,
  \dodoi{10.3847/2041-8213/aac7c9}

\bibitem[{{Galv{\'a}n-Madrid} {et~al.}(2018){Galv{\'a}n-Madrid}, {Liu},
  {Izquierdo}, {Miotello}, {Zhao}, {Carrasco-Gonz{\'a}lez}, {Lizano}, \&
  {Rodr{\'\i}guez}}]{Galvan-Madrid2018}
{Galv{\'a}n-Madrid}, R., {Liu}, H.~B., {Izquierdo}, A.~F., {et~al.} 2018, \apj,
  868, 39, \dodoi{10.3847/1538-4357/aae779}

\bibitem[{{Gammie}(2001)}]{Gammie2001}
{Gammie}, C.~F. 2001, \apj, 553, 174, \dodoi{10.1086/320631}

\bibitem[{{Goldreich} {et~al.}(1986){Goldreich}, {Goodman}, \&
  {Narayan}}]{Goldreich1986}
{Goldreich}, P., {Goodman}, J., \& {Narayan}, R. 1986, \mnras, 221, 339,
  \dodoi{10.1093/mnras/221.2.339}

\bibitem[{{Gundlach} {et~al.}(2018){Gundlach}, {Schmidt}, {Kreuzig},
  {Bischoff}, {Rezaei}, {Kothe}, {Blum}, {Grzesik}, \& {Stoll}}]{Gundlach2018}
{Gundlach}, B., {Schmidt}, K.~P., {Kreuzig}, C., {et~al.} 2018, \mnras, 479,
  1273, \dodoi{10.1093/mnras/sty1550}

\bibitem[{{Harsono} {et~al.}(2018){Harsono}, {Bjerkeli}, {van der Wiel},
  {Ramsey}, {Maud}, {Kristensen}, \& {J{\o}rgensen}}]{Harsono2018}
{Harsono}, D., {Bjerkeli}, P., {van der Wiel}, M. H.~D., {et~al.} 2018, Nature
  Astronomy, 2, 646, \dodoi{10.1038/s41550-018-0497-x}

\bibitem[{{Harsono} {et~al.}(2021){Harsono}, {van der Wiel}, {Bjerkeli},
  {Ramsey}, {Calcutt}, {Kristensen}, \& {J{\o}rgensen}}]{Harsono2021}
{Harsono}, D., {van der Wiel}, M.~H.~D., {Bjerkeli}, P., {et~al.} 2021, \aap,
  646, A72, \dodoi{10.1051/0004-6361/202038697}

\bibitem[{{Hogg} {et~al.}(2010){Hogg}, {Bovy}, \& {Lang}}]{Hogg2010}
{Hogg}, D.~W., {Bovy}, J., \& {Lang}, D. 2010, arXiv e-prints, arXiv:1008.4686,
  \dodoi{10.48550/arXiv.1008.4686}

\bibitem[{{Huang} {et~al.}(2018){Huang}, {Andrews}, {Dullemond}, {Isella},
  {P{\'e}rez}, {Guzm{\'a}n}, {{\"O}berg}, {Zhu}, {Zhang}, {Bai}, {Benisty},
  {Birnstiel}, {Carpenter}, {Hughes}, {Ricci}, {Weaver}, \&
  {Wilner}}]{dsharpII}
{Huang}, J., {Andrews}, S.~M., {Dullemond}, C.~P., {et~al.} 2018, \apjl, 869,
  L42, \dodoi{10.3847/2041-8213/aaf740}

\bibitem[{{Kimura} {et~al.}(2015){Kimura}, {Wada}, {Senshu}, \&
  {Kobayashi}}]{Kimura2015}
{Kimura}, H., {Wada}, K., {Senshu}, H., \& {Kobayashi}, H. 2015, \apj, 812, 67,
  \dodoi{10.1088/0004-637X/812/1/67}

\bibitem[{{Klahr} {et~al.}(2018){Klahr}, {Pfeil}, \& {Schreiber}}]{Klahr2018}
{Klahr}, H., {Pfeil}, T., \& {Schreiber}, A. 2018, in Handbook of Exoplanets,
  ed. H.~J. {Deeg} \& J.~A. {Belmonte}, 138,
  \dodoi{10.1007/978-3-319-55333-7_138}

\bibitem[{{Ko} {et~al.}(2020){Ko}, {Liu}, {Lai}, {Ching}, {Rao}, \&
  {Girart}}]{Ko2020ApJ...889..172K}
{Ko}, C.-L., {Liu}, H.~B., {Lai}, S.-P., {et~al.} 2020, \apj, 889, 172,
  \dodoi{10.3847/1538-4357/ab5e79}

\bibitem[{{Kratter} \& {Lodato}(2016)}]{KratterLodato2016}
{Kratter}, K., \& {Lodato}, G. 2016, \araa, 54, 271,
  \dodoi{10.1146/annurev-astro-081915-023307}

\bibitem[{{Kristensen} {et~al.}(2012){Kristensen}, {van Dishoeck}, {Bergin},
  {Visser}, {Y{\i}ld{\i}z}, {San Jose-Garcia}, {J{\o}rgensen}, {Herczeg},
  {Johnstone}, {Wampfler}, {Benz}, {Bruderer}, {Cabrit}, {Caselli}, {Doty},
  {Harsono}, {Herpin}, {Hogerheijde}, {Karska}, {van Kempen}, {Liseau},
  {Nisini}, {Tafalla}, {van der Tak}, \& {Wyrowski}}]{Kristensen2012}
{Kristensen}, L.~E., {van Dishoeck}, E.~F., {Bergin}, E.~A., {et~al.} 2012,
  \aap, 542, A8, \dodoi{10.1051/0004-6361/201118146}

\bibitem[{{Lee} {et~al.}(2020){Lee}, {Li}, \& {Turner}}]{Lee2020}
{Lee}, C.-F., {Li}, Z.-Y., \& {Turner}, N.~J. 2020, Nature Astronomy, 4, 142,
  \dodoi{10.1038/s41550-019-0905-x}

\bibitem[{{Lesur} {et~al.}(2022){Lesur}, {Ercolano}, {Flock}, {Lin}, {Yang},
  {Barranco}, {Benitez-Llambay}, {Goodman}, {Johansen}, {Klahr}, {Laibe},
  {Lyra}, {Marcus}, {Nelson}, {Squire}, {Simon}, {Turner}, {Umurhan}, \&
  {Youdin}}]{Lesur2022}
{Lesur}, G., {Ercolano}, B., {Flock}, M., {et~al.} 2022, arXiv e-prints,
  arXiv:2203.09821, \dodoi{10.48550/arXiv.2203.09821}

\bibitem[{{Li} {et~al.}(2017){Li}, {Liu}, {Hasegawa}, \& {Hirano}}]{Li2017}
{Li}, J. I.-H., {Liu}, H.~B., {Hasegawa}, Y., \& {Hirano}, N. 2017, \apj, 840,
  72, \dodoi{10.3847/1538-4357/aa6f04}

\bibitem[{{Lin} \& {Pringle}(1987)}]{LinPringle1987}
{Lin}, D.~N.~C., \& {Pringle}, J.~E. 1987, \mnras, 225, 607,
  \dodoi{10.1093/mnras/225.3.607}

\bibitem[{{Liu}(2019)}]{Liu2019}
{Liu}, H.~B. 2019, \apjl, 877, L22, \dodoi{10.3847/2041-8213/ab1f8e}

\bibitem[{{Liu}(2021)}]{Liu2021}
---. 2021, \apj, 914, 25, \dodoi{10.3847/1538-4357/abf8b6}

\bibitem[{{Liu} {et~al.}(2021){Liu}, {Tsai}, {Chen}, {Liu}, {Zhang}, {Ma},
  {Elbakyan}, {Green}, {Hales}, {Liu}, {Takami}, {P{\'e}rez}, {Vorobyov}, \&
  {Yang}}]{Liu2021b}
{Liu}, H.~B., {Tsai}, A.-L., {Chen}, W.~P., {et~al.} 2021, \apj, 923, 270,
  \dodoi{10.3847/1538-4357/ac31b9}

\bibitem[{{Lodato} \& {Rice}(2004)}]{LodatoRice2004}
{Lodato}, G., \& {Rice}, W.~K.~M. 2004, \mnras, 351, 630,
  \dodoi{10.1111/j.1365-2966.2004.07811.x}

\bibitem[{{Lodato} {et~al.}(2023){Lodato}, {Rampinelli}, {Viscardi},
  {Longarini}, {Izquierdo}, {Paneque-Carre{\~n}o}, {Testi}, {Facchini},
  {Miotello}, {Veronesi}, \& {Hall}}]{Lodato2023}
{Lodato}, G., {Rampinelli}, L., {Viscardi}, E., {et~al.} 2023, \mnras, 518,
  4481, \dodoi{10.1093/mnras/stac3223}

\bibitem[{{McClure} {et~al.}(2016){McClure}, {Bergin}, {Cleeves}, {van
  Dishoeck}, {Blake}, {Evans}, {Green}, {Henning}, {{\"O}berg}, {Pontoppidan},
  \& {Salyk}}]{McClure2016}
{McClure}, M.~K., {Bergin}, E.~A., {Cleeves}, L.~I., {et~al.} 2016, \apj, 831,
  167, \dodoi{10.3847/0004-637X/831/2/167}

\bibitem[{{Musiolik} \& {Wurm}(2019)}]{Musiolik2019}
{Musiolik}, G., \& {Wurm}, G. 2019, \apj, 873, 58,
  \dodoi{10.3847/1538-4357/ab0428}

\bibitem[{{Nakatani} {et~al.}(2020){Nakatani}, {Liu}, {Ohashi}, {Zhang},
  {Hanawa}, {Chandler}, {Oya}, \& {Sakai}}]{Nakatani2020}
{Nakatani}, R., {Liu}, H.~B., {Ohashi}, S., {et~al.} 2020, \apjl, 895, L2,
  \dodoi{10.3847/2041-8213/ab8eaa}

\bibitem[{{Ohashi} {et~al.}(2022){Ohashi}, {Nakatani}, {Liu}, {Kobayashi},
  {Zhang}, {Hanawa}, \& {Sakai}}]{Ohashi2022}
{Ohashi}, S., {Nakatani}, R., {Liu}, H.~B., {et~al.} 2022, \apj, 934, 163,
  \dodoi{10.3847/1538-4357/ac794e}

\bibitem[{{Ormel} \& {Cuzzi}(2007)}]{OrmelCuzzi2007}
{Ormel}, C.~W., \& {Cuzzi}, J.~N. 2007, \aap, 466, 413,
  \dodoi{10.1051/0004-6361:20066899}

\bibitem[{{Paneque-Carre{\~n}o} {et~al.}(2021){Paneque-Carre{\~n}o},
  {P{\'e}rez}, {Benisty}, {Hall}, {Veronesi}, {Lodato}, {Sierra}, {Carpenter},
  {Andrews}, {Bae}, {Henning}, {Kwon}, {Linz}, {Loinard}, {Pinte}, {Ricci},
  {Tazzari}, {Testi}, \& {Wilner}}]{Paneque2021}
{Paneque-Carre{\~n}o}, T., {P{\'e}rez}, L.~M., {Benisty}, M., {et~al.} 2021,
  \apj, 914, 88, \dodoi{10.3847/1538-4357/abf243}

\bibitem[{{Pillich} {et~al.}(2021){Pillich}, {Bogdan}, {Landers}, {Wurm}, \&
  {Wende}}]{Pillich2021}
{Pillich}, C., {Bogdan}, T., {Landers}, J., {Wurm}, G., \& {Wende}, H. 2021,
  \aap, 652, A106, \dodoi{10.1051/0004-6361/202140601}

\bibitem[{{Pineda} {et~al.}(2022){Pineda}, {Arzoumanian}, {Andr{\'e}},
  {Friesen}, {Zavagno}, {Clarke}, {Inoue}, {Chen}, {Lee}, {Soler}, \&
  {Kuffmeier}}]{Pineda2022}
{Pineda}, J.~E., {Arzoumanian}, D., {Andr{\'e}}, P., {et~al.} 2022, arXiv
  e-prints, arXiv:2205.03935, \dodoi{10.48550/arXiv.2205.03935}

\bibitem[{{Rafikov}(2015)}]{Rafikov2015}
{Rafikov}, R.~R. 2015, \apj, 804, 62, \dodoi{10.1088/0004-637X/804/1/62}

\bibitem[{{Riols} {et~al.}(2017){Riols}, {Latter}, \&
  {Paardekooper}}]{RiolsLatter2017}
{Riols}, A., {Latter}, H., \& {Paardekooper}, S.~J. 2017, \mnras, 471, 317,
  \dodoi{10.1093/mnras/stx1548}

\bibitem[{{Segura-Cox} {et~al.}(2020){Segura-Cox}, {Schmiedeke}, {Pineda},
  {Stephens}, {Fern{\'a}ndez-L{\'o}pez}, {Looney}, {Caselli}, {Li}, {Mundy},
  {Kwon}, \& {Harris}}]{Segura-Cox2020}
{Segura-Cox}, D.~M., {Schmiedeke}, A., {Pineda}, J.~E., {et~al.} 2020, \nat,
  586, 228, \dodoi{10.1038/s41586-020-2779-6}

\bibitem[{{Sheehan} {et~al.}(2020){Sheehan}, {Tobin}, {Federman}, {Megeath}, \&
  {Looney}}]{Sheehan2020}
{Sheehan}, P.~D., {Tobin}, J.~J., {Federman}, S., {Megeath}, S.~T., \&
  {Looney}, L.~W. 2020, \apj, 902, 141, \dodoi{10.3847/1538-4357/abbad5}

\bibitem[{{Sheehan} {et~al.}(2022){Sheehan}, {Tobin}, {Looney}, \&
  {Megeath}}]{Sheehan2022}
{Sheehan}, P.~D., {Tobin}, J.~J., {Looney}, L.~W., \& {Megeath}, S.~T. 2022,
  \apj, 929, 76, \dodoi{10.3847/1538-4357/ac574d}

\bibitem[{{Steinpilz} {et~al.}(2019){Steinpilz}, {Teiser}, \&
  {Wurm}}]{Steinpilz2019}
{Steinpilz}, T., {Teiser}, J., \& {Wurm}, G. 2019, \apj, 874, 60,
  \dodoi{10.3847/1538-4357/ab07bb}

\bibitem[{{Tazzari} {et~al.}(2021){Tazzari}, {Clarke}, {Testi}, {Williams},
  {Facchini}, {Manara}, {Natta}, \& {Rosotti}}]{Tazzari2021tension}
{Tazzari}, M., {Clarke}, C.~J., {Testi}, L., {et~al.} 2021, \mnras, 506, 2804,
  \dodoi{10.1093/mnras/stab1808}

\bibitem[{{Terry} {et~al.}(2022){Terry}, {Hall}, {Longarini}, {Lodato}, {Toci},
  {Veronesi}, {Paneque-Carre{\~n}o}, \& {Pinte}}]{Terry2022}
{Terry}, J.~P., {Hall}, C., {Longarini}, C., {et~al.} 2022, \mnras, 510, 1671,
  \dodoi{10.1093/mnras/stab3513}

\bibitem[{{Tobin} {et~al.}(2020){Tobin}, {Sheehan}, {Megeath},
  {D{\'\i}az-Rodr{\'\i}guez}, {Offner}, {Murillo}, {van 't Hoff}, {van
  Dishoeck}, {Osorio}, {Anglada}, {Furlan}, {Stutz}, {Reynolds}, {Karnath},
  {Fischer}, {Persson}, {Looney}, {Li}, {Stephens}, {Chandler}, {Cox},
  {Dunham}, {Tychoniec}, {Kama}, {Kratter}, {Kounkel}, {Mazur}, {Maud},
  {Patel}, {Perez}, {Sadavoy}, {Segura-Cox}, {Sharma}, {Stephenson}, {Watson},
  \& {Wyrowski}}]{Tobin2020}
{Tobin}, J.~J., {Sheehan}, P.~D., {Megeath}, S.~T., {et~al.} 2020, \apj, 890,
  130, \dodoi{10.3847/1538-4357/ab6f64}

\bibitem[{{Tripathi} {et~al.}(2017){Tripathi}, {Andrews}, {Birnstiel}, \&
  {Wilner}}]{Tripathi2017}
{Tripathi}, A., {Andrews}, S.~M., {Birnstiel}, T., \& {Wilner}, D.~J. 2017,
  \apj, 845, 44, \dodoi{10.3847/1538-4357/aa7c62}

\bibitem[{{Tripathi} {et~al.}(2018){Tripathi}, {Andrews}, {Birnstiel},
  {Chandler}, {Isella}, {P{\'e}rez}, {Harris}, {Ricci}, {Wilner}, {Carpenter},
  {Calvet}, {Corder}, {Deller}, {Dullemond}, {Greaves}, {Henning}, {Kwon},
  {Lazio}, {Linz}, \& {Testi}}]{Tripathi2018}
{Tripathi}, A., {Andrews}, S.~M., {Birnstiel}, T., {et~al.} 2018, \apj, 861,
  64, \dodoi{10.3847/1538-4357/aac5d6}

\bibitem[{{Tsukamoto} {et~al.}(2022){Tsukamoto}, {Maury}, {Commer{\c{c}}on},
  {Alves}, {Cox}, {Sakai}, {Ray}, {Zhao}, \& {Machida}}]{Tsukamoto2022}
{Tsukamoto}, Y., {Maury}, A., {Commer{\c{c}}on}, B., {et~al.} 2022, arXiv
  e-prints, arXiv:2209.13765, \dodoi{10.48550/arXiv.2209.13765}

\bibitem[{{Veronesi} {et~al.}(2021){Veronesi}, {Paneque-Carre{\~n}o}, {Lodato},
  {Testi}, {P{\'e}rez}, {Bertin}, \& {Hall}}]{Veronesi2021}
{Veronesi}, B., {Paneque-Carre{\~n}o}, T., {Lodato}, G., {et~al.} 2021, \apjl,
  914, L27, \dodoi{10.3847/2041-8213/abfe6a}

\bibitem[{{Vorobyov} \& {Basu}(2007)}]{VorobyovBasu2007}
{Vorobyov}, E.~I., \& {Basu}, S. 2007, \mnras, 381, 1009,
  \dodoi{10.1111/j.1365-2966.2007.12321.x}

\bibitem[{{Vorobyov} \& {Basu}(2010)}]{VorobyovBasu2010}
---. 2010, \apj, 719, 1896, \dodoi{10.1088/0004-637X/719/2/1896}

\bibitem[{{Vorobyov} \& {Elbakyan}(2019)}]{Vorobyov2019}
{Vorobyov}, E.~I., \& {Elbakyan}, V.~G. 2019, \aap, 631, A1,
  \dodoi{10.1051/0004-6361/201936132}

\bibitem[{{Vorobyov} {et~al.}(2023){Vorobyov}, {Elbakyan}, {Johansen},
  {Lambrechts}, {Skliarevskii}, \& {Stoyanovskaya}}]{Vorobyov2023}
{Vorobyov}, E.~I., {Elbakyan}, V.~G., {Johansen}, A., {et~al.} 2023, \aap, 670,
  A81, \dodoi{10.1051/0004-6361/202244500}

\bibitem[{{Wada} {et~al.}(2009){Wada}, {Tanaka}, {Suyama}, {Kimura}, \&
  {Yamamoto}}]{Wada2009}
{Wada}, K., {Tanaka}, H., {Suyama}, T., {Kimura}, H., \& {Yamamoto}, T. 2009,
  \apj, 702, 1490, \dodoi{10.1088/0004-637X/702/2/1490}

\bibitem[{{Windmark} {et~al.}(2012){Windmark}, {Birnstiel}, {G{\"u}ttler},
  {Blum}, {Dullemond}, \& {Henning}}]{Windmark2012}
{Windmark}, F., {Birnstiel}, T., {G{\"u}ttler}, C., {et~al.} 2012, \aap, 540,
  A73, \dodoi{10.1051/0004-6361/201118475}

\bibitem[{{Xu}(2022)}]{Xu22}
{Xu}, W. 2022, \apj, 934, 156, \dodoi{10.3847/1538-4357/ac7b94}

\bibitem[{{Xu} \& {Armitage}(2023)}]{XA23}
{Xu}, W., \& {Armitage}, P.~J. 2023, \apj, 946, 94,
  \dodoi{10.3847/1538-4357/acb7e5}

\bibitem[{{Xu} \& {Kunz}(2021{\natexlab{a}})}]{XK21a}
{Xu}, W., \& {Kunz}, M.~W. 2021{\natexlab{a}}, \mnras, 502, 4911,
  \dodoi{10.1093/mnras/stab314}

\bibitem[{{Xu} \& {Kunz}(2021{\natexlab{b}})}]{XK21b}
---. 2021{\natexlab{b}}, \mnras, 508, 2142, \dodoi{10.1093/mnras/stab2715}

\bibitem[{{Yamamuro} {et~al.}(2023){Yamamuro}, {Tanaka}, \&
  {Okuzumi}}]{Yamamuro2023}
{Yamamuro}, R., {Tanaka}, K. E.~I., \& {Okuzumi}, S. 2023, \apj, 949, 29,
  \dodoi{10.3847/1538-4357/acc52f}

\bibitem[{{Youdin} \& {Goodman}(2005)}]{YoudinGoodman2005}
{Youdin}, A.~N., \& {Goodman}, J. 2005, \apj, 620, 459, \dodoi{10.1086/426895}

\bibitem[{{Zamponi} {et~al.}(2021){Zamponi}, {Maureira}, {Zhao}, {Liu}, {Ilee},
  {Forgan}, \& {Caselli}}]{Zamponi2021}
{Zamponi}, J., {Maureira}, M.~J., {Zhao}, B., {et~al.} 2021, \mnras, 508, 2583,
  \dodoi{10.1093/mnras/stab2657}

\bibitem[{{Zhang} {et~al.}(2018){Zhang}, {Zhu}, {Huang}, {Guzm{\'a}n},
  {Andrews}, {Birnstiel}, {Dullemond}, {Carpenter}, {Isella}, {P{\'e}rez},
  {Benisty}, {Wilner}, {Baruteau}, {Bai}, \& {Ricci}}]{dsharpVII}
{Zhang}, S., {Zhu}, Z., {Huang}, J., {et~al.} 2018, \apjl, 869, L47,
  \dodoi{10.3847/2041-8213/aaf744}

\bibitem[{{Zhu} {et~al.}(2010){Zhu}, {Hartmann}, {Gammie}, {Book}, {Simon}, \&
  {Engelhard}}]{Zhu2010}
{Zhu}, Z., {Hartmann}, L., {Gammie}, C.~F., {et~al.} 2010, \apj, 713, 1134,
  \dodoi{10.1088/0004-637X/713/2/1134}

\bibitem[{{Zhu} {et~al.}(2019){Zhu}, {Zhang}, {Jiang}, {Kataoka}, {Birnstiel},
  {Dullemond}, {Andrews}, {Huang}, {P{\'e}rez}, {Carpenter}, {Bai}, {Wilner},
  \& {Ricci}}]{Zhu2019}
{Zhu}, Z., {Zhang}, S., {Jiang}, Y.-F., {et~al.} 2019, \apjl, 877, L18,
  \dodoi{10.3847/2041-8213/ab1f8c}

\end{thebibliography}
\bibliographystyle{aasjournal}

\end{document}